\documentclass[lettersize,journal]{IEEEtran}
\usepackage{amsmath,amsfonts}
\usepackage{amsthm} %NEWLY ADDED SALIM FOR PROOF
\allowdisplaybreaks
\usepackage{algorithmic}
\usepackage{algorithm}
\usepackage{array}
\usepackage[caption=false,font=normalsize,labelfont=sf,textfont=sf]{subfig}
\usepackage{textcomp}
\usepackage{stfloats}
\usepackage{url}
\usepackage{verbatim}
\usepackage{graphicx}
\usepackage{cite}
\newcommand{\PLOS}{\mathop{\Pr}\nolimits_{\mathrm{LOS}}}
\newcommand{\PNLOS}{\mathop{\Pr}\nolimits_{\mathrm{NLOS}}}
\newcommand{\PLMAX}{\mathrm{PL_{\max}}}

\newtheorem{theorem}{Theorem}

\begin{document}

\title{Integrating UAV-Enabled Base Stations in 3D Networks: QoS-Aware Joint Fronthaul and Backhaul Design}

\author{Salim~Janji,~\IEEEmembership{Graduate Student Member,~IEEE,}
        Piotr~Wawrzyniak,
        Piotr~Formanowicz,
        Adrian~Kliks,~\IEEEmembership{Senior Member,~IEEE}\thanks{This manuscript is a preprint of a paper currently under review at IEEE Journal on Selected Areas in Communications (IEEE JSAC). Copyright © 2024 IEEE. Personal use of this material is permitted. Permission from IEEE must be obtained for all other uses, in any current or future media, including reprinting/republishing this material for advertising or promotional purposes, creating new collective works, for resale or redistribution to servers or lists, or reuse of any copyrighted component of this work in other works.}}

\markboth{IEEE Journal on Selected Areas in Communications}%
{Janji \MakeLowercase{\textit{et al.}}}


\maketitle

\begin{abstract}
The emerging concept of 3D networks, integrating terrestrial, aerial, and space layers, introduces a novel and complex structure characterized by stations relaying backhaul loads through point-to-point wireless links, forming a wireless 3D backhaul mesh. A key challenge is the strategic placement of aerial platform such as drone base stations (DBSs), considering the locations and service demands of ground nodes and the connectivity to backhaul gateway nodes for core network access. This paper addresses these complexities with a two-fold approach: a novel Agglomerative Hierarchical Clustering (HC) algorithm that optimizes DBS locations to satisfy minimum backhaul adjacency and maximum fronthaul coverage radius requirements; and a Genetic Algorithm (GA) that designs backhaul connections to satisfy the cumulative load across the network and maximize the throughput margin which translates to network resilience to increasing demands. Our results showcase the effectiveness of these algorithms against benchline schemes, offering insights into the operational dynamics of these novel 3D networks.
\end{abstract}


\section{Introduction}
\IEEEPARstart{D}{rone} base stations (DBSs) are emerging as a popular solution within beyond fifth-generation (B5G) and sixth-generation (6G) networks for providing wireless connectivity in different scenarios such as covering disaster-struck areas, improving coverage and capacity of existing terrestrial networks, and supplying uncovered areas \cite{tutorial_mozaffari, tutorial_zeng_1,tutorial_zeng_2}. DBSs are faster and cheaper to deploy when compared to terrestrial base stations. Also, they offer more flexibility due to their three-dimensional mobility which also increases the probability of achieving line-of-sight (LOS) links with ground users due to their higher altitude. Meanwhile, owing to their varying features such as size, weight, payload capacity, energy source, etc., unmanned aerial vehicles (UAVs) \footnote{In this paper, the term "Drone" is assumed to encompass all types of UAVs, including multi-rotor drones, fixed-wing drones, and balloons, which are considered functionally similar for the purposes of telecommunications optimization.} offer a wide range of capabilities which can serve diverse deployment schemes with different mission times, coverage areas, and location densities. For example, to maximize spectrum reuse and LOS probability in an urban area, dense deployment of multiple DBSs can be employed with transceivers operating in millimeter wave (mmWave) \cite{mmwave_clustering} or sub-6 GHz \cite{monte_carlo_mobiquitous} frequencies. Otherwise, to provide coverage to large areas, scarce deployment of high-altitude drones or platforms such as balloons with sub-6 GHz transceivers has also received attention. \textit{Loon} project \cite{loon_fso} which leveraged stratospheric balloons and \textit{Free Basics} project \cite{meta_free_basics} using drones are two major examples of such deployments that aim to cover underserved regions across the globe.

\begin{figure}[!htb]
\centering
  \includegraphics[width=0.47\textwidth]{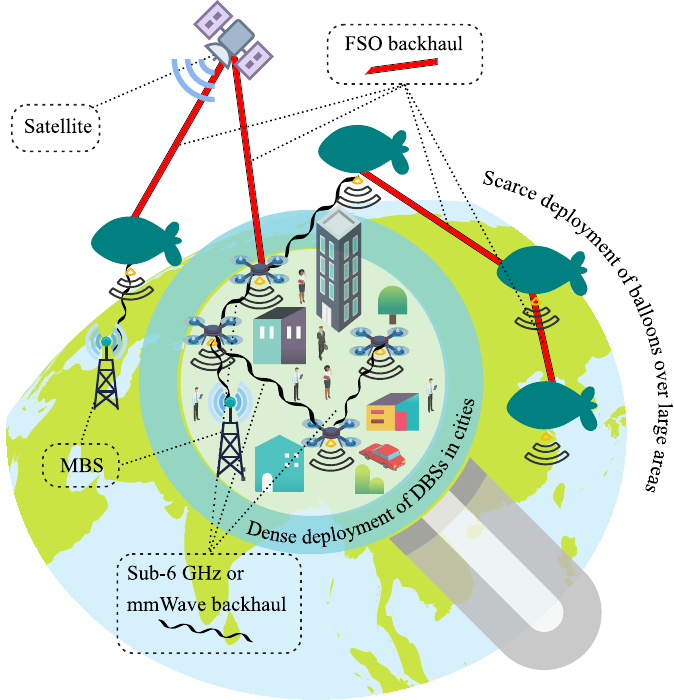}
  \caption{Illustration of diverse UAV-based 3D network configurations for B5G and 6G wireless communication. Low-altitude DBSs facilitate dense network deployments in urban settings to enhance spectrum reuse and LOS connectivity. High-altitude platforms, such as balloons, extend coverage across vast areas. Backhaul solutions include inter-UAV links and connections to terrestrial base stations or satellites, utilizing FSO, cmWave, and mmWave technologies for robust wireless communication.}
  \label{fig:scenario}
\end{figure}

Regardless of the scenario and deployment parameters, DBSs serving as access points require backhaul to the core network. Unlike their terrestrial counterparts, DBSs must use wireless links that can leverage free-space optical (FSO) communication \cite{fso_backhaul_fronthaul}, cmWave and mmWave channels \cite{mmwave_backhaul}, or sub-6 GHz technologies. As described in \cite{tutorial_mozaffari}, a DBS can establish backhaul connectivity through connecting to a terrestrial macro base station (MBS) or a satellite. Alternatively, DBSs can be interconnected with each other or with additional aerial platforms, constructing what is currently referred to as a three-dimensional (3D) network\cite{3d_network}. Then, one or more stations that are connected to the core can serve as gateways to the established backhaul mesh network such that a DBS can reach the core network either through direct connection or via other DBSs. Fig.~\ref{fig:scenario} illustrates the different scenarios and backhaul options described. 

% Examples of such networks include those formed by low-power internet of things (IoT) nodes (e.g., home appliances) which leverage Zigbee, Z-Wave, and Thread protocols.

The depicted configuration, involving the interconnection of multiple nodes, gives rise to a \textit{wireless mesh} network \cite{backhaul_survey_2010} or \textit{wireless multi-hop} network \cite{backhaul_survey_2022}. These structures are increasingly prominent in contemporary telecommunication systems across various scales. Within the mesh, cells can also form a self-organizing network (SON) in terms of topology, wherein the network proactively establishes alternative paths to redirect traffic and maintain uninterrupted connectivity in the event of node failures or disconnections. Besides 3D networks, a similar backhaul connectivity problem is presented by ultra-dense networks (UDNs) that are key enablers of the main 5G objectives such as high data rates, low latency, and massive connectivity \cite{udn_survey}. For backhaul, UDNs may utilize different types of wired links (e.g., fiber optic), point-to-point (ptp) microWave or mmWave, satellite, or sub-6 GHz links\cite{backhaul_survey_2022}. Network operators planning those links consider a variety of factors including the feasibility of wired connections, clear LOS for microWave or mmWave frequencies, interference, costs, etc. 

Precise construction of backhaul links is crucial to avoid potential bottlenecks in network performance, encompassing factors such as latency, throughput, and security. When deploying DBSs, utilizing mmWave and FSO technologies offers favorable solutions for backhaul provisioning. Examples of projects which utilized an FSO backhaul network in their designs are presented in \cite{fso_survey_acq}. Leveraging these technologies provides dedicated and enhanced throughput capabilities within a wider available bandwidth, surpassing the capabilities of sub-6 GHz technologies \cite{backhaul_survey_2022}. Furthermore, they offer heightened security due to the channel's directivity, making it exceedingly challenging for interceptions to go undetected. Additionally, FSO transceivers can be placed at distances of up to 10 km or more, contingent upon atmospheric conditions and the specific layer of the atmosphere in which they operate. Additionally, to increase the robustness of the network, and since each technology is affected by different weather factors, extensive research has been conducted on the hybrid deployment of both FSO and mmWave systems\cite{backhaul_survey_2022, fso_mmwave_hybrid}. 

Multi-hop backhaul mesh networks formed with DBSs leveraging mmWave or (and) FSO technologies present a novel scenario. Firstly, due to their altitude, DBSs operating in the same region would have a clear LOS path between each other in most cases which results in a large number of possible connections between different pairs. Secondly, the channel conditions of the ptp links are of variable nature due to the DBSs' mobility and higher sensitivity to atmospheric factors due to the larger distance between nodes. The mobility factor may also implicate changes in the network topology which in turn requires severing links and configuring new links between different pairs of DBSs. Unlike network cells with omnidirectional propagation, only a DBSs pair with transceivers aligned can establish a link. Given their locations, the alignment process can leverage various tracking and pointing mechanisms that are already in use for networks involving UAVs, high-speed trains, nanosatellites, and terrestrial base stations \cite{fso_survey_acq}. However, this implies that the DBSs cannot perform neighbor discovery by themselves similarly to e.g., IoT nodes, but it's rather a central controller that determines the pairs of connected DBSs and feeds the proper location information. Finally, the limited payload weight and energy storage capacity of UAVs impose restrictions on the number of transceivers they can accommodate, which constrains the number of simultaneous links that a DBS can establish with other stations.

This novel network architecture introduces a new optimization problem which involves placing the DBSs and selecting a single backhaul interconnection configuration out of many possibilities given the backhaul load produced by the served traffic of each DBS, its transmission parameters, and the constraint on the number of backhaul connections it can have. Another factor is the reliability of the network which we consider in two ways: a) how much further unexpected throughput can the backhaul network serve? and, b) in case of a backhaul link failure, how many other neighbors does a disconnected DBS have for re-establishing connectivity?

\subsection{Related Work}

In \cite{fronthaul_backhaul_single_hop}, the authors develop a solution for the deployment and resource management of DBSs considering both fronthaul and backhaul constraints, focusing on the direct connectivity of each DBS to a macro base station for backhaul. Their approach involves a joint optimization algorithm that handles the dual aspects of user association and UAV placement, leveraging a branch-and-bound method to minimize duality gaps. Contrastingly, our work elaborates on a wider scenario where DBSs can relay backhaul traffic between each other, forming a 3D network mesh. This introduces an additional layer to the problem formulation and the placement strategy, as DBSs must be positioned not only concerning ground users but also relative to other DBSs to maintain network connectivity and optimize the wireless backhaul mesh network.
In \cite{tethered_backhaul}, DBSs can relay backhaul traffic to each other and terrestrial stations while providing coverage to ground nodes (GNs). However, DBSs backhaul links are assumed tethered and their candidate locations defined a priori. Moreover, the solution state space is limited as follows. For each GN, a number of pre-calculated shortest paths are considered to undergo a selection process via a deep learning technique to maximize network efficiency and coverage.
Our work closely resembles the study in \cite{same_backhaul}, where DBSs are deployed to assist small cells (SCs) and establish a backhaul mesh network among themselves. In this mesh, the backhaul capacity of a path to a gateway is determined by its weakest link. Through a game-theoretic framework, each DBS aims to optimize its backhaul throughput by choosing its connections with other DBSs or MBSs. However, their model does not restrict the number of backhaul connections for DBSs, nor does it consider the cumulative load of serviced GNs along these paths. Similarly, \cite{similar_work} addresses the challenge of positioning DBSs to guarantee complete coverage for GNs, ensuring that all DBSs maintain connectivity, whether direct or through intermediary DBSs, yet it leaves the assignment of gateway DBSs to the core network unspecified. Their approach confines DBS locations to predefined grid points and overlooks the actual backhaul throughput demands stemming from the GNs and relaying other DBSs' backhaul data. Also, the throughput of a backhaul link is neglected such that only the existence of a connection is considered. In contrast, our work accounts for the achievable rate of each DBS backhaul. We also enforce that each backhaul should meet the demands of the GNs within its associated DBS cluster, along with the load from other DBSs relaying data to a designated MBS gateway via this DBS, as depicted in Fig. \ref{fig:example}.

\subsection{Main Contributions}
An offline algorithm is introduced consisting of an agglomerative hierarchical clustering (HC) solution for placing the DBSs and a genetic algorithm (GA) for selecting the backhaul interconnection configuration. The key contributions of this article are as follows.

\begin{enumerate}
\item Given the locations of core gateway nodes, as well as the location and rate requirements of GNs, the number of backhaul transceivers per DBS, and the maximum distance for backhaul links, we explore the placement of DBSs and their backhaul interconnections by formulating the primary optimization problem, which is subsequently divided into two subproblems after proving its NP-hardness.
\item We propose an agglomerative HC algorithm for clustering GNs to ensure that the resulting minimal number of DBS clusters and their locations meet the minimum required number of neighbors for establishing backhaul connectivity for each DBS. Increasing the number of neighbors results in increasing the reliability of the backhaul network and the quality of the solutions obtained in the second subproblem.
\item We formulate the Drone Network Problem (DNP) which involves selecting the backhaul interconnection edges that satisfy each cluster's throughput demand while maximizing surplus throughput.
\item After proving its NP-Hardness, we introduce a novel GA algorithm that efficiently solves the DNP while increasing the level of backhaul throughput redundancy.
\end{enumerate}

\subsection{Paper Notations and Organization}
The paper starts with an overview of the system model, detailing the access and backhaul layers in Section~\ref{sec:system_model}. It then outlines the main optimization challenge, breaking it down into DBS placement and backhaul network optimization in Section~\ref{sec:problem_formulation}. Following this, Sections \ref{sec:hc} and \ref{sec:ga} introduce the HC and GA approaches devised to address these issues. Section~\ref{sec:sim} presents the simulation outcomes, and Section~\ref{sec:conclusion} provides concluding remarks. A summary of principal notations used throughout the paper and their definitions can be found in Table~\ref{tab:notations}.

%TwoColumn
\begin{table*}[ht]
\centering
\caption{MAJOR NOTATIONS USED IN THIS PAPER}
\label{tab:notations}
\begin{tabular}{|l|l|l|l|}
\hline
\multicolumn{4}{|c|}{\textbf{Symbol and Description}} \\
\hline
$U$ The number of GNs & $\mathcal{V}$ The set of BSs & $h$ DBS height & $d_\text{max}$ Maximum distance of backhaul link \\
$M$ The number of DBSs & $\mathcal{M}$ The set of DBSs &$\mathcal{S}^p$ Sequence denoting backhaul path $p$ & $R(i,l)$ Backhaul throughput between BSs $i$ and $l$\\
$B$ The number of MBSs & $\mathcal{B}$ The set of MBSs & $G(j)$ Total GNs load for cluster $j$ &$\mathcal{S}^p_r$ Element $r$ of backhaul sequence $p$\\
$R_n$ Load rate of GN $n$ & $\boldsymbol u_n$ Location of GN $n$ &$N_\text{B}$ Minimum number of DBS neighbors&$W(j,n)$ Association indicator for DBS $j$ and GN $n$\\

\hline
\end{tabular}
\end{table*}

% \begin{table*}[ht]
% \centering
% \caption{MAJOR NOTATIONS USED IN THIS PAPER}
% \label{tab:notations}
% \begin{tabular}{|l|l|l|}
% \hline
% \multicolumn{3}{|c|}{\textbf{Symbol and Description}} \\
% \hline
% $U$ The number of GNs & $\mathcal{V}$ The set of BSs & $d_\text{max}$ Maximum distance of backhaul link \\
% $M$ The number of DBSs & $\mathcal{M}$ The set of DBSs &$\mathcal{S}^p$ Sequence denoting backhaul path $p$\\
% $B$ The number of MBSs & $\mathcal{B}$ The set of MBSs & $G(j)$ Total GNs load for cluster $j$\\
% $R_n$ Load rate of GN $n$ & $\boldsymbol u_n$ Location of GN $n$ &$N_\text{B}$ Minimum number of DBS neighbors\\
% $h$ DBS height & $R(i,l)$ Backhaul throughput between BSs $i$ and $l$ & $\mathcal{S}^p_r$ Element $r$ of backhaul sequence $p$\\
% \hline
% \end{tabular}
% \end{table*}

\section{System Model}\label{sec:system_model}
We consider a $10$x$10$ km rectangular coverage area, within which ground nodes are spatially distributed following a Poisson cluster process (PCP). This modeling approach effectively captures the spatial arrangement of various GNs, including SCs, macrocell base stations (MBSs), and ground users in heterogeneous cellular networks (HetNets) \cite{pcp_user_modeling}. There are $U$ ground nodes in total, and each GN $n \in \{1, 2,..,U\}$ is defined by its 2D location,\footnote{We presuppose a uniform ground elevation for all terrestrial nodes, a justifiable approximation considering that DBSs are presumed to operate at relatively elevated altitudes.} $\boldsymbol u_n = (x_n, y_n)$, and its expected throughput load $R_n$. To cover these nodes, $M$ DBSs are deployed, thus forming a set $\mathcal{M}$ with cardinality $M$. Similarly, the location of DBS $j$ is denoted by $\mathbf{m}_j = (x_j, y_j, h)$. We assume an equal altitude for all DBSs, which not only simplifies the problem but also constitutes a pragmatic assumption allowing reduced complexity of the onboard backhaul hardware (i.e., circumvent scenarios in which the body of a DBS might obstruct transmission). Finally, let $\mathcal{B}$ represent the set comprising the $B$ gateway access points available for gateway connections of the backhaul mesh network. These access points can be comprised of other aerial platforms, satellites, or MBSs. Without loss of generality, we consider MBSs in this work. These MBSs are defined by their positions given by $\mathbf{b}_k = (x_k, y_k, h_k)$, where $k$ ranges from $1$ to $B$. Combining the sets of DBSs, $\mathcal{M}$, and MBSs, $\mathcal{B}$, we obtain the comprehensive BS set $\mathcal{V} = \mathcal{B} \cup \mathcal{M} = \{\mathcal{B}_1, ..., \mathcal{B}_B, \mathcal{M}_1, .., \mathcal{M}_M\}$ with a total of $B+M$ elements.

\subsection{GN-DBS Path Loss Model (Access Layer)}
The 3D Euclidean distance between points indexed by $ j$ and $n $ is given as 
\begin{equation}\label{eqn:distance_3d}
d(j, n) = \sqrt{d_{\text{2d}}^2(j, n) + (h_j - h_n)^2},
\end{equation}
where 
\begin{equation}\label{eqn:distance_2d}
d_{\text{2d}}(j, n) = \sqrt{(x_j-x_n)^2 + (y_j-y_n)^2}
\end{equation}
represents the 2D distance on the ground plane. This formula calculates distances between various pairs of points, such as two GNs, two BSs, or a GN and a BS, in our study. Specifically, in this section, $d(j, n)$ refers to the distance between DBS $ j$ and GN $ n $. In this work, we utilize the prevalent simplified path-loss (PL) model, as outlined in \cite{optimalLapAltitude}, which articulates the probability of a channel maintaining LOS as opposed with non-line-of-sight (NLOS) status as
\begin{equation}
  \PLOS(j, n) = \frac{1}{1+\alpha \times e^{-\beta (\omega - \alpha)}},\label{eqn:plos}
\end{equation}
where $\alpha$ and $\beta$ are environment-dependent variables and $\omega = \arctan(\frac{h_j}{d_{j,n}})$. Let \(\eta_{\mathrm{LOS}}\) and \(\eta_{\mathrm{NLOS}}\) respectively denote the anticipated PL for LOS and NLOS links in dB. The overall expected PL in dB can be formulated as
%TwoColumn
\begin{multline}\label{eqn:expectedPathLoss}
  L^{\mathrm{[dB]}}(j, n) = 20 \log\left(\frac{4\pi f_cd(j, n)}{c}\right) +\\ \PLOS(j, n) \times \eta_{\mathrm{LOS}} + \PLOS(j, n) \times \eta_{\mathrm{NLOS}},
\end{multline}
% \begin{equation}\label{eqn:expectedPathLoss}
%   L^{\mathrm{[dB]}}(j, n) = 20 \log\left(\frac{4\pi f_cd(j, n)}{c}\right) + \PLOS(j, n) \times \eta_{\mathrm{LOS}} + \PNLOS(j, n) \times \eta_{\mathrm{NLOS}},
% \end{equation}
where $f_c$ and $c$ symbolize the carrier frequency and the speed of light, respectively. Let $\PLMAX$ represent the maximal permissible PL, determined by specific transmission power and a designated signal-to-noise ratio (SNR). As delineated in \cite{optimalLapAltitude}, this parameter allows for the derivation of the coverage disk radius $R_\text{A}$, adhering to the $\PLMAX$ constraint, for a fixed DBS height value. Hence, a GN, $n$, falls under the service of DBS $j$ if it lies within its coverage disk\footnote{In this study, we set aside the effects of shadowing, fading, and interference, concentrating primarily on coverage via placement and maintaining adequate backhaul throughput. Such assumptions can be substantiated through proper scheduling strategies or by segmenting clusters into smaller sub-clusters, with lower-altitude DBSs reutilizing the spectrum.}, and this is mapped accordingly by the binary variable 
\begin{equation}\label{eqn:gn_map}
 W(j,n) = 
 \begin{cases}
      w(j,n) & \text{if $d_{\text{2d}}(j,n) < R_\text{A}$},\\
      0 & \text{otherwise}.
    \end{cases}
\end{equation}
where $w(j,n)$ is a binary variable determined by the association mechanism which ensures that no GN is served by more than one DBS in case of overlapping coverage disks (i.e., $\sum_{n=1}^U =1$).
\subsection{FSO Channel Model (Backhaul Layer)}
To model the backhaul channel for simulations, we employ FSO links. However, it is imperative to note that our algorithms are applicable for all wireless backhauling schemes such as mmWave; the only difference being in the resulting throughput value between BSs pairs. We leverage the model presented in \cite{fso_model}, in which the FSO channel gain is given by
\begin{equation}\label{eqn:fsoGain}
    h_\text{FSO} = \eta h_p h_a h_g,
\end{equation}
where $\eta$ is the photo-detector responsivity and $h_p$, $h_a$, and $h_g$ are the atmospheric loss, atmospheric turbulence induced fading, and geometric and misalignment losses (GML), respectively. Denote the distance in meters for the FSO link between BSs $i$ and $l$, with indices $i, l \in \{1, ..., B+M\}$, as $d(i, l)$. Then, the atmospheric loss, $h_p$, can be expressed as 
\begin{equation}\label{eqn:atmLoss}
 h_p = 10^{-\kappa \frac{d}{10}},
 \end{equation}
where BS indices are omitted in equations \eqref{eqn:atmLoss}-\eqref{eqn:FsoRate} for simplicity. The coefficient $\kappa$ $[m^{-1}]$ represents a weather-dependent attenuation coefficient, accounting for various weather conditions, including clear air and varying fog intensities. Concerning the atmospheric turbulence, \(h_a\), which is induced by temperature and atmospheric pressure inconsistencies across the link, the authors in \cite{fso_model} utilize its expected value, assuming its impact is insubstantial relative to the GML. This can be ensured by employing adaptive optics (AO) technology that usually leverages deformable mirrors in the FSO receivers which is proven to mitigate atmospheric-turbulence-induced losses \cite{fso_survey_acq}. Then, adopting the same assumption implies \(h_a = \mathbb{E}\{h_a\} = 1\). Regarding the GML, the authors presuppose a link distance spanning several hundred meters, sufficiently exceeding the FSO beam footprint misalignment (i.e., the distance between the incident beam's center on the receiver's lens and the lens's mid-point). This assumption is applicable to our system, where the link distance may even extend into the kilometer range. Therefore, following the same assumptions for each FSO backhaul link, we first determine the beam width in meters at distance $d$ as
\begin{equation}
    \omega_d = \omega_0 \sqrt{1 + \left(1+\frac{2\omega_0^2}{\rho^2(d)}\right)\left(\frac{\lambda d}{\pi\omega^2_0}\right)^2},
\end{equation}
where \(\omega_0\) signifies the beam waist radius, and the coherence length \(\rho(d)\) is given by \(\rho(d)=\left(0.55C^2_nk^2L\right)^{-3/5}\), dependent on \(k=\frac{2\pi}{\lambda}\) with \(\lambda\) as the optical wavelength, and \(C_n^2 \approx C_0^2 \exp(-\frac{h}{100})\) is the index of refraction structure parameter at DBS height \(h\), while \(C_0^2=1.7 \times 10^{-14}\) represents the ground nominal value of refractive index. To proceed further, we highlight that in this work we have assumed that the beam is perpendicular on the receiver's lens plane. This assumption is valid for a majority of conventional FSO transceiver architectures mounted on UAVs. These often utilize mechanisms such as gimbals, mirrors, and rotating heads, or a combination thereof, to ensure alignment \cite{fso_survey_acq}. Then given a lens radius $r_0$, we obtain the coefficient $v_1=\frac{r_0}{\omega_L}\sqrt{\frac{\pi}{2}}$ which is then used to determine the maximum fraction of optical power captured by the receiver lens denoted by
\begin{equation}
    A_0=\left(\text{erf}(v_1)\right)^2.
\end{equation}
Finally, the GML is given by
\begin{equation}\label{eqn:gml}
    h_g \approx A_0 \exp \left(-\frac{2u^2}{\zeta\omega^2_d}y\right),
\end{equation}
where $u$ represents the misalignment distance (in meters) between the centers of the beam footprint and the receiver's lens. The parameter $\zeta$ is selected from the interval $[\zeta_1,\zeta_2]$, providing bounds on the GML loss. In this work, $\zeta$ is chosen as the arithmetic mean, i.e., $\zeta=\frac{\zeta_1 + \zeta_2}{2}$. In \cite{fso_model}, the authors prioritize the statistical variations of $u$ as the primary source of GML loss fluctuations. They model the position and orientation perturbations of the DBS - resulting from atmospheric air fluctuations and DBS vibrations - as independent distributed Gaussian RVs. Consequently, definitions are given as \(\lambda_1=\sigma_y^2+d^2\sigma_\theta^2\) and \(\lambda_2=\sigma_z^2+d^2\sigma_\phi^2\), where \(\sigma_y\) and \(\sigma_z\) denote the variances of location fluctuations perpendicular to the beam and along the height axis, respectively, while \(\sigma_\theta\) and \(\sigma_\phi\) indicate variances in orientation fluctuations due to rotations around vertical and horizontal axes, respectively. Then, the rate of an FSO link is given by
\begin{equation}\label{eqn:FsoRate}
    R_\text{FSO}= \frac{1}{2}\log_2\left(\frac{e}{2\pi}\eta^2h_p^2
\bar{\gamma}A_0^2\right) - \frac{2}{\zeta\omega^2_d \ln(2)} (\lambda_1 + \lambda_2),
\end{equation}
with \(\bar{\gamma} = \frac{P_\text{FSO}^2}{\sigma_n^2}\) representing the transmit SNR, defined in terms of the FSO transmission power, \(P_\text{FSO}\), and the noise at the receiver, \(\sigma_n\). Finally, by constraining the maximum FSO link distance to $d_\text{max}$ the achievable throughput of an FSO link between BSs $i$ and $l$ is given by the following equation where we have returned the indices. 
\begin{equation}\label{eqn:FsoRateFinal}
 R(i,l) = 
 \begin{cases}
      R_\text{FSO}(i,l) & \text{if } d(i, l) < d_\text{max},\\
      0 & \text{otherwise}.
    \end{cases}
\end{equation}

\subsection{Backhaul Network}\label{sec:backhaul_network}
\begin{figure*}[!htb]
\centering
  \includegraphics[width=0.65\linewidth, trim={0.3cm 0.2cm 0cm 0.5cm},clip]{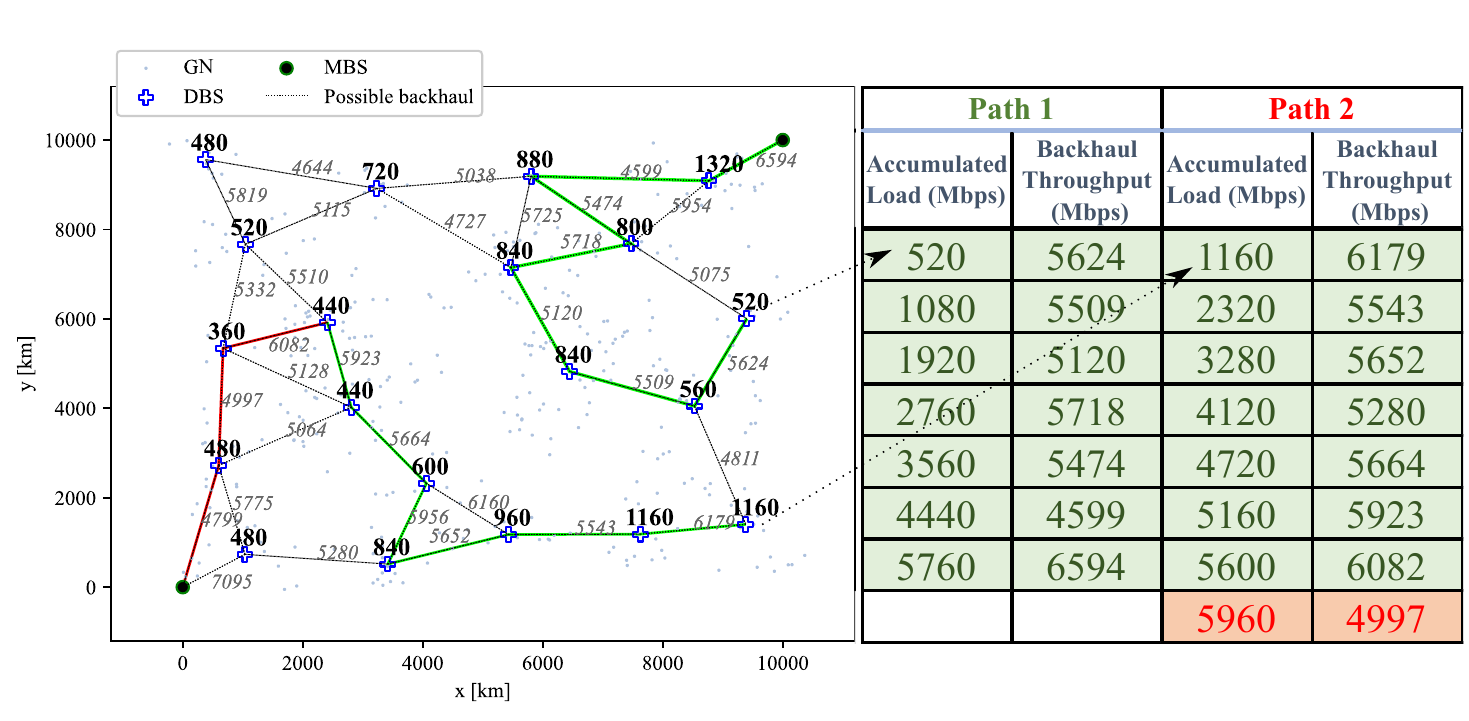}
  \caption{Visualization of the backhaul optimization, with pre-determined DBS locations and GNs loads. Potential backhaul links are shown with throughput values. Each DBS's total GNs load is labeled. Two sample connectivity paths are highlighted: green lines signify sufficient backhaul capacity, and red indicates congestion with respective accumulated loads.}
  \label{fig:example}
\end{figure*}

The network architecture is represented by a graph $G = (\mathcal{V}, \mathcal{E})$, where $\mathcal{V}$ denotes the set of BSs previously defined, and $\mathcal{E} = \{\{v_i, v_j\} | v_i, v_j \in \mathcal{V}, i \neq j\}$ comprises the potential connections between them. These connections are weighted according to the FSO throughputs, as calculated by \eqref{eqn:FsoRate}. Additionally, each node is tagged with the load from its serviced GN cluster. Using \eqref{eqn:gn_map}, this load can be obtained for DBS $j$ by 
\begin{equation}
    G(j) = \sum_{n=1}^U W(j,n) R_n.
\end{equation}
Fig.~\ref{fig:example} presents an example of this network with DBSs, two MBSs, and the possible interconnections between them adhering to the $ d_\text{max}$ constraint. The figure also shows throughput and total load annotations on top of the FSO links and DBSs. To clarify the problem further, we also plot two examples of backhaul paths extending from a DBS endpoint to an MBS gateway. Along each path, each DBS, $j$, contributes the load in Mbps of its associated GNs, $G(j)$, to its forward backhaul link which should withstand this load in addition to the loads of previous DBSs, if any. The green path successfully aggregates loads across DBS relays, while the red indicates an overload scenario where the backhaul inter-connectivity fails in managing an accumulated load of 5960 Mbps just two hops away from the MBS gateway. 

\section{Problem Formulation}\label{sec:problem_formulation}
We now present the main optimization problem and show how it corresponds to an NP-hard problem, and subsequently partition it into two distinct subproblems that we addressed separately.

\subsection{Main Problem}\label{subsection:opt_problem_main}
With the positions $\boldsymbol{u}_n$ and rates $R_n$ of each GN, indexed by $n \in {1, ..., U}$, and the predefined backhaul distance limit $d_\text{max}$, we aim to determine the set of DBSs, $\mathcal{M}$. This includes determining their number and specific locations. Subsequently, this allows for the calculation of the potential backhaul connection throughputs between pairs of BSs, denoted as $R(i,l)$, and the associated load $G(j)$ for each GN cluster served by a DBS $j$, as derived from \eqref{eqn:FsoRateFinal} and \eqref{eqn:gn_map}. Another decision variable is determining a set of $B$ sequences, $\mathcal{S} = \{\mathcal{S}^1, ...,\mathcal{S}^B\}$ each denoting a backhaul path with at least two BS indices. Each non-empty sequence begins with an a DBS, and ends with an MBS concluding the backhaul path. The optimization problem is then given by
\begin{subequations}\label{eqn:optimization0}
\begin{align}
    &\min_{\mathbf{m}_j, \mathcal{S}} \;\; M \label{eqn:opt0_obj}\\
    \text{s.t.\;\;} & \sum_{j=1}^M W(j,n) \geq 1\;\;\;\;\;\; \forall n \in \{1,..,U\}\label{eqn:opt0_cond0}\\
     & \bigcup^B_{p=1} \mathcal{S}^p \supset \mathcal{M}\label{eqn:opt0_cond1}\\
    & \mathcal{S}^p \cap \mathcal{S}^{p'} = \{\phi\} \;\;\;\;\;\; \forall p,  p' \in \{1,..,B\} \;\;, \;\; p\neq p'\label{eqn:opt0_cond2}\\
    & \mathcal{S}^p_{|\mathcal{S}^p|} \in  \mathcal{B} \;\;\;\;\;\; \forall p \in \{1,..,B\}\label{eqn:opt0_cond3}\\
    &  R(\mathcal{S}^p_q, \mathcal{S}^p_{q+1}) \geq \sum_{r=1}^{t} G(\mathcal{S}^p_r) \label{eqn:opt0_cond4}\\
    &\;\;\;\;\;\;\;\;\;\;\;\;\;\;\;\;\;\;\;\;\;\;\;\;\forall p \in \{1,..,B\}, \;\; \forall q \in \{1,..,|\mathcal{S}^p| - 1\}\notag\\
    & \mathcal{S}^p_i, \mathcal{S}^p_j \in \mathcal{S}^p, \; i \neq j \Rightarrow \mathcal{S}^p_i \neq \mathcal{S}^p_j \label{eqn:opt0_cond5}\\ 
    &\;\;\;\;\;\;\;\;\;\;\;\;\;\;\;\;\;\;\;\;\;\;\;\;\forall p \in \{1,..,B\}, \;\; \forall i,j \in \{1,..,|\mathcal{S}^p|\}\notag
\end{align}
\end{subequations}
Objective \eqref{eqn:opt0_obj} seeks to minimize the number of DBSs deployed while the constraints entail the following.
\begin{itemize}
    \item \eqref{eqn:opt0_cond0}: Each GNs is covered with at least one DBS.
    \item \eqref{eqn:opt0_cond1}: All DBSs are considered within the resulting sequences.
    \item \eqref{eqn:opt0_cond2}: Ensures exclusivity of each DBS to one path.
    \item \eqref{eqn:opt0_cond3}: Imposes that the concluding BS in every path is an MBS.
    \item \eqref{eqn:opt0_cond4}: Given $|\mathcal{S}^p|$ as the cardinality of sequence $p$, this constraint guarantees each backhaul link can cater to the accumulated demand from prior DBSs.
    \item \eqref{eqn:opt0_cond5}: Implies that the BSs within each set are distinct to ensure that no DBS has more than two FSO connections.
\end{itemize}
Our problem is similar to the geometric set cover (GSC) problem, which is a classical example of NP-hard problems. In GSC, one aims to encompass a specified set of points in $\mathbb{R}^2$ using the least number of predefined geometric entities, such as circles with a designated radius. Aligning with the approach in \cite{similar_work}, if we ease the backhaul restrictions, setting both $d_\text{max}$ and $ R(i,l) $ for all $i,l\in \{1, ..., B+M\}$ to infinity, our problem reverts to a canonical GSC challenge. Here, the goal becomes covering a set of points in $\mathbb{R}^2$ using the fewest circles of radius $R$, analogous to the coverage circles of the DBSs. This proves that the problem described in \eqref{eqn:opt0_obj} is NP-hard. To proceed with solving the problem we propose subdividing it into two subproblems as follows.

\subsection{Number of DBSs and their Locations}
The first subdivision of the problem deals with determining the number and locations of DBSs, taking into account the locations and rates of GNs. After this, we focus on establishing the backhaul connections. When selecting DBS locations, it's essential that each DBS is located within a distance of $d_\text{max}$ from another DBS. This ensures that every DBS has a potential backhaul connection, regardless of throughput needs. We extend this constraint by imposing that every DBS must be within an $d_\text{max}$ distance of at least $N_\text{B}$ other DBSs. By increasing the arbitrary value of $N_\text{B}$, the network is more densely connected which allows more solutions of backhaul connections in the next subproblem. Therefore, given 
\begin{equation}\label{eqn:fso_map}
 W^\text{FSO}(i, l) = 
 \begin{cases}
      1 & \text{if } d(i, l) < d_\text{max},\\
      0 & \text{otherwise},
    \end{cases}
\end{equation}
as an indicator function which checks whether a BS pair is within $d_\text{max}$ of each other, this problem is formally expressed as \eqref{eqn:opt0_obj}, adhering to the constraints \eqref{eqn:opt0_cond0} and
\begin{equation}\label{eqn:neighbor_constraint}
    \deg(v_j) = \sum_{i=1, i \neq j}^{B+M} W^\text{FSO}(i, j) \geq N_\text{B} \;\;\;\; \forall j \in \{1,..., B+M\}.
\end{equation}
Clearly, this still resembles a GSC problem with the added constraint about the number of nearby disks for each disk and therefore it's also NP-hard.

\subsection{Backhaul Optimization - Drone Network Problem (DNP)}\label{subsec:dnp}
As for the second subproblem illustrated in Fig.~\ref{fig:example} that highlights the challenges in selecting optimal backhaul links while catering to the accumulated throughput demands, we start by introducing another optimization objective that was not included in \eqref{eqn:optimization0}. Beyond merely meeting the backhaul demands, there's also an intent to maximize the extra throughput in each backhaul path, anticipating future changes in ground network needs. This problem can be formulated as follows. Given the set of BSs, $\mathcal{V}$ with determined locations, potential backhaul connections' achievable throughput between each BS pair as $R(i,l)$, and the load $G(j)$ of each GNs cluster, the goal is to find a set of $B$ sequences as we described earlier in Subsection \ref{subsection:opt_problem_main}. By considering the secondary objective of surplus throughput, these sequences should satisfy

\begin{subequations}\label{eqn:optimization2}
\begin{align}
    \max_{\mathcal{S}^1, ...,\mathcal{S}^B} & \prod_{p=1}^B   \left( \sum_{r=1}^{|\mathcal{S}^p| - 1} R(\mathcal{S}^p_r, \mathcal{S}^p_{r+1}) - G(\mathcal{S}^p_r)\right)\label{eqn:opt1}\\
    \text{s.t.\;\;} & \eqref{eqn:opt0_cond1}, \eqref{eqn:opt0_cond2}, \eqref{eqn:opt0_cond3}, \eqref{eqn:opt0_cond4}, \text{ and } \eqref{eqn:opt0_cond5}
\end{align}
\end{subequations}

This DNP can be formulated as a graph problem involving the previously defined graph $G = (\mathcal{V}, \mathcal{E})$ (see \ref{sec:backhaul_network}). To formally define the problem, we consider the following DNP.

\noindent
{\sc Drone Network Problem (DNP)}\\
\noindent
{\sc Instance}:\\
Undirected graph $G = (\mathcal{V}, \mathcal{E})$, where $\mathcal{V} = \mathcal{B} \cup \mathcal{M}$, positive integer $G(j)$ for each vertex $v_j \in \mathcal{M}$, positive integer $R(i,l)$ for each edge $\{v_i,v_l\}\in \mathcal{E}$.\\
%Below is already defined SJ
% ($\mathcal{B}$ represents the set of base stations, $\mathcal{M}$ denotes the set of drones, $G(j)$ corresponds to the data bandwidth requirement per drone and $R(i,l)$ represents the throughput of the link between drones or between drones and base stations.)
\noindent
{\sc Answer}:\\
Subgraph $G_S=(\mathcal{V}_S,\mathcal{E}_S)$ of graph $G$ such that $\mathcal{V}_S = \mathcal{V}$, $\mathcal{E}_S\subseteq \mathcal{E}$, $\forall_{v_i\in \mathcal{M}}\deg(v_i)\leq 2$, $\forall_{v_i\in \mathcal{B}}\deg(v_i)\leq 1$ and in $G_S$ there can be distinguished set of vertex sequences $\mathcal{S} = \{\mathcal{S}^1, ...,\mathcal{S}^B\}$ such that every vertex $v_i\in \mathcal{V}_S$ is an element of at most one sequence $\mathcal{S}^p \in \mathcal{S}$, and every vertex from set $\mathcal{M}$ belongs to exactly one sequence $\mathcal{S}^p \in \mathcal{S}$. Moreover, for each sequence $\mathcal{S}^p=(v_{p_1},v_{p_2},\ldots,v_{p_l})\in \mathcal{S}$ the following conditions are satisfied: $\forall_{i=1,2,\ldots,l-1}v_{p_i}\in \mathcal{M}$, $v_{p_l}\in \mathcal{B}$ and $\forall_{i=1,2,\ldots,l-1}R(p_i,p_{i+1})\geq \sum_{k=1}^{i}G(p_k)$ (see \eqref{eqn:opt0_cond4}).

The DNP is computationally intractable, which is stated in the following theorem.
\begin{theorem}
The Drone Network Problem is NP-hard.
\end{theorem}
\begin{proof}
NP-hardness of DNP will be proved by showing that its decision counterpart is NP-complete. A problem is NP-complete if it belongs to class NP and some other NP-complete problem can be polynomially transformed to it.\cite{garey1979computers}. In the decision counterpart of the DNP, the instance is the same as in DNP, but the answer is YES if a subgraph $G_S$ exists, otherwise the answer is NO. As a known NP-complete problem, we have chosen the widely known Hamiltonian Path problem (HPP)\cite{garey1979computers}, which can be defined as:\\
\noindent
{\sc Hamiltonian Path Problem (HPP)}\\
\noindent
{\sc Instance}:\\
Undirected graph $G_h = (V_h, E_h)$\\
\noindent
{\sc Answer}:\\
YES, if in $G_h$ there is a Hamiltonian path, i.e., path $S_h = (v_1, v_2, ..., v_{|V_h|})$ visiting every vertex from $V_h$ exactly once; NO, otherwise.\\
Given an instance of the HPP, the following instance of the DNP is created:
$G = (\mathcal{V}, \mathcal{E})$, where $\mathcal{V} = V_h \cup \{b\}$, $\mathcal{M} = V_h$, $\mathcal{B} = \{b\}$, $\mathcal{E} = E_h \cup \{\{v_1,b\}, \{v_2,b\}, \dots, \{v_{|V_h|},b\}\}$, $\forall_{v_j \in \mathcal{M}}{G(j) = 1}$ and $\forall_{\{v_i,v_l\} \in \mathcal{E}}{R(i,l) = |V_h|}$

$\Rightarrow$ If for the HPP the answer is YES, the answer is also YES for the instance of the decision version of the DNP. Easily, we obtain a single path $\mathcal{S}^h = (v_1, v_2, ..., v_{|V_h|}, b)$, which is the solution subgraph $G_S$ for the DNP. The condition \eqref{eqn:opt0_cond4} is also fulfilled, because it sums $|V_h|$ values of $1$ for $G(i)$, when each link throughput $R(i,l)$ is equal to $|V_h|$.

$\Leftarrow$ If we have an answer YES for the instance of the decision version of the DNP, a graph $G_S$ exists that can be partitioned into a set of vertex sequences $\mathcal{S} = \{\mathcal{S}^1, ..., \mathcal{S}^B\}$. As we have added only one vertex $b \in \mathcal{B}$, in $\mathcal{S}$ we have a single sequence $\mathcal{S}^b$ ending with vertex $b$, i.e., $\mathcal{S}^b = (v_{i_1}, v_{i_2}, ..., v_{i_{|V_h|}}, b)$. After removing the last node $b$, we obtain the Hamiltonian path $S_h = (v_{i_1}, v_{i_2}, ..., v_{i_{|V_h|}})$ and the answer for the instance of the HPP is YES.

We have demonstrated that the decision version of the DNP is NP-complete. Consequently, this implies that the DNP (its search version) is NP-hard.
\end{proof}

Since the DNP is NP-hard, it strongly suggests that attempting to find an exact solution using an exact algorithm may be impractical given its potentially exponential or greater complexity. Therefore, instead of pursuing an exact solution, we find an approximate solution using a GA which will efficiently guide us toward an optimal or near-optimal solution.

\section{Hierarchical Clustering (HC) for Selecting Number of DBSs and Their Locations}\label{sec:hc}
In this work, we utilize agglomerative HC, a bottom-up method that starts with each data point as an independent cluster and progressively merges them based on similarity or distance metrics. This approach, with our proposed modifications, ensures the number of neighbors constraint, $N_B$, as specified in $\eqref{eqn:neighbor_constraint}$, and guarantees that the clustered GNs fall within the serving DBS's coverage area, denoted as $R_A$. We adapt the unweighted pair group method using centroids (UPGMC)\cite{hierc_clustering} within the agglomerative HC framework to specify the quantity and positions of DBSs. These adaptations introduce additional merging criteria to meet our constraints. Beginning with the 2D location of each GN as a separate cluster, the algorithm merges clusters based on Euclidean distance until a single cluster is formed or a predetermined stopping criterion is met.

Firstly, a dissimilarity matrix of size $U\times U$ is constructed, capturing the 2D distances between all GN pairs using \eqref{eqn:distance_2d}. The same formula is also used to compute the distance between the centers of any two clusters, $\mathcal{C}_a$ and $\mathcal{C}_{a'}$, where both $a$ and $a'$ belong to ${1,...,U}$. We define the symmetric dissimilarity matrix at iteration $t$, $\boldsymbol{D}^t(\mathcal{U})$, as

\begin{equation}\label{eqn:dissimilarity_matrix}
    \boldsymbol{D}^0(\mathcal{U})= \begin{bmatrix}
0 & d_{\text{2d}}(1, 2) &  d_{\text{2d}}(1, 3) & \cdots & d_{\text{2d}}(1, U)\\
d_{\text{2d}}(2, 1) & 0 &  d_{\text{2d}}(2, 3) & \cdots & d_{\text{2d}}(2, U)\\
 \vdots &  \vdots &   \vdots & \ddots & \vdots \\
 d_{\text{2d}}(U, 1) &  d_{\text{2d}}(U, 2) &  d_{\text{2d}}(U, 3) & \cdots & 0
\end{bmatrix}.
\end{equation}
Here, $D_{a,b}$ at the $a$th row and $b$th column indicates $d_{\text{2d}}(a, b)$. In each iteration, $t$, the algorithm first identifies a subset, $\mathcal{P}$, of elements within the matrix, rather than the entire matrix. The criteria for determining this subset will be defined in the subsequent sections. Then, the pair of clusters achieving the smallest distance corresponding to the element
\begin{equation}\label{eqn:minimum_distance}
D^t_{a,b} = \min_{p,q\in \mathcal{P}}D^t_{p,q}
\end{equation}
are merged to form a new cluster, necessitating an update to the distances between other clusters and this newly-formed cluster, thereby generating $\boldsymbol{D}^{t+1}(\mathcal{U})$. If the resultant cluster is expressed as $\mathcal{C}_q$, emerging from the union of clusters $\mathcal{C}_a$ and $\mathcal{C}_b$, then $\mathcal{C}_q = \mathcal{C}_a \cup \mathcal{C}_b$. For computational efficiency within the UPGMC algorithm, the distance between another cluster, $\mathcal{C}_s$, and the newly formed $\mathcal{C}_q$ in the subsequent dissimilarity matrix is \cite{hierc_clustering}
% TwoColumn
\begin{multline}\label{eqn:upgmc_update}
    d_{\text{2d}}(q, s) = \frac{n_a}{n_a + n_b} d_{\text{2d}}(a, s) + \frac{n_b}{n_a + n_b} d_{\text{2d}}(b, s)\\ - \frac{n_a n_b}{(n_a + n_b)^2} d_{\text{2d}}(a, b)
\end{multline}
% \begin{equation}\label{eqn:upgmc_update}
%     d_{\text{2d}}(q, s) = \frac{n_a}{n_a + n_b} d_{\text{2d}}(a, s) + \frac{n_b}{n_a + n_b} d_{\text{2d}}(b, s)\\ - \frac{n_a n_b}{(n_a + n_b)^2} d_{\text{2d}}(a, b),
% \end{equation}
where the number of elements in $\mathcal{C}_a$ and $\mathcal{C}_b$ are represented by $n_a$ and $n_b$ respectively. Utilizing the UPGMC version of HC clustering through \eqref{eqn:upgmc_update} ensures that the updated dissimilarity matrix reflects the squared Euclidean distance between the cluster pair means\cite{hierc_clustering}. Thus, the distance between cluster centers, equivalent to the distance between DBSs as given by \eqref{eqn:distance_2d}, corresponds to the values in the dissimilarity matrix.

In our adaptation of UPGMC, we derive a binary matrix, $ \hat{\boldsymbol{D}^t}$, using the mask function $F$: $\mathbb{R}^{m \times n} \rightarrow \mathbb{R}^{m \times n}$. This function renders the new matrix elements, $\hat{D^t}_{a,b}$, as
\begin{equation}\label{eqn:degrees_matrix}
     \hat{D}^t_{a,b} = F(D_{a,b}) = \begin{cases}
      1 & \text{if } d_{\text{2d}}(a, b) < d_\text{max} \text{ and } a\neq b,\\
      0 & \text{otherwise}.
    \end{cases}
\end{equation}
After computing $\hat{\boldsymbol{D}^t}$, a column summation yields a vector detailing the count of neighboring clusters each cluster possesses within the range of $d_\text{max}$. This vector is expressed as
\begin{equation}\label{eqn:degrees_vector}
     \boldsymbol{C}^0 =\begin{bmatrix}
        \sum^U_{a'=1} \hat{D}_{1,a'}\\
        \vdots\\
        \sum^U_{a'=1} \hat{D}_{U,a'}
    \end{bmatrix}.
\end{equation}
Let's assume that a DBS is placed at each cluster center at iteration $t$ for a total of $U - t$ clusters, it can be readily observed that 
\begin{equation}\label{eqn:degrees_vector_equal_neighbor}
     \boldsymbol{C}^t =\begin{bmatrix}
        \sum^{U - t}_{a'=1} \hat{D}_{1,a'}\\
        \vdots\\
        \sum^{U - t}_{a'=1} \hat{D}_{U,a'}
    \end{bmatrix} = \begin{bmatrix}
        \sum_{a'=1, a' \neq 1}^{{U - t}} W^\text{FSO}(1, a')\\
        \vdots\\
        \sum^{U - t}_{a'=1, a' \neq U - t} W^\text{FSO}(U - t, a')
    \end{bmatrix}.
\end{equation}
By employing \eqref{eqn:degrees_matrix} and \eqref{eqn:degrees_vector}, we can ensure that no cluster has fewer neighbors than $N_\text{B}$ by doing the following. Let's consider an example, if $\mathcal{C}_c$ is a cluster with only one neighbor, $\mathcal{C}_a$, (meaning $C_c=1$) and $N_\text{B}=1$, any clustering of $\mathcal{C}_a$ may alter its central location. Merging it with a different cluster than $\mathcal{C}_c$ may relocate $\mathcal{C}_a$ out of $\mathcal{C}_c$'s vicinity, violating the $N_\text{B}$ constraint. To circumvent this, before merging clusters $\mathcal{C}_a$ and $\mathcal{C}_b$, we impose the condition that every non-zero element in the vector
\begin{equation}
     \boldsymbol{V} = \left(\hat{D}^t_{:,a} \oplus \hat{D}^t_{:,b}\right) \circ \boldsymbol{C}^t
\end{equation}
exceeds $N_\text{B}$, where $\circ$ represents the Hadamard product and $\oplus$ signifies the binary OR operation. Specifically, the following condition must be satisfied:
\begin{equation}\label{eqn:hierc_clus_cond_1}
\forall V_i \in \boldsymbol{V}, i \neq a, i \neq b : V_i > 0 \implies V_i > N_\text{B}.
\end{equation}
If this condition isn't met, we skip the current cluster pairing and look for the next pair that aligns with \eqref{eqn:minimum_distance} for potential merging. The modification we have described ensures that each formed cluster has at least $N_\text{B}$ neighbors. 

Furthermore, to guarantee that all GNs covered by a single DBS lie within a radius of $R_\text{A}$, we employ the complete link (CL) HC algorithm's update equation \cite{hierc_clustering}. Like \eqref{eqn:upgmc_update}, the distance between a cluster $\mathcal{C}_s$ and a newly merged cluster $\mathcal{C}_q$, derived from the combination of clusters $\mathcal{C}_a$ and $\mathcal{C}_b$, is expressed as:
\begin{equation}\label{eqn:complete_link_update}
    d_{\text{2d}}(q, s) = \frac{1}{2} \left( d_{\text{2d}}(a, s) + d_{\text{2d}}(b, s) + | d_{\text{2d}}(a, s) - d_{\text{2d}}(b, s) | \right).
\end{equation}
Starting with the dissimilarity matrix $\boldsymbol{D}^0$, we use \eqref{eqn:complete_link_update} to generate an updated dissimilarity matrix, $\boldsymbol{R}^t$. As indicated in \cite{hierc_clustering}, the entry $R^t_{a,b}$ represents the largest distance between any two points within clusters $\mathcal{C}_a$ and $\mathcal{C}_b$.
\begin{theorem}
Combining clusters $\mathcal{C}_a$ and $\mathcal{C}_b$ at iteration \( t \), where the associated dissimilarity in $\boldsymbol{R}^t$ adheres to
\begin{equation}\label{eqn:hierc_clus_cond_2}
    R^t_{a,b} \leq R_\text{A}
\end{equation}
ensures that every GN within the merged cluster lies inside a radius of $R_\text{A}$ from its DBS, irrespective of the cluster's centroid.
\end{theorem}
\begin{proof}
Let $\mathcal{C}_c = \mathcal{C}_a \cup \mathcal{C}_b$ be resulting merged set of points with cardinality $n_c$, and let the squared distance between its two farthest points, $\mathcal{C}_{c,A}$ and $\mathcal{C}_{c,B}$, be
\begin{equation}\label{eqn:max_distance_pts}
     R^2_\text{A} =  (x_a - x_b)^2 + (y_a - y_b)^2.
\end{equation}
The centroid of $\mathcal{C}_c$ is 
\begin{equation}
    \bar{P}_{\mathcal{C}_c} = \left(\frac{\sum^{n_c}_{i=1}x_i}{n_c}, \frac{\sum^{n_c}_{i=1}y_i}{n_c}\right),
\end{equation}
where $(x_i, y_i)$ are the coordinates of point $i$ in $\mathcal{C}_c$. The squared distance from any point $\mathcal{C}_{c,P}$ to the centroid is
\begin{equation}\label{eqn:centroid_to_point}
    d_{\text{2d}}^2(\mathcal{C}_{c,P}, \bar{P}_{\mathcal{C}_c}) = \left(x_p - \frac{\sum^{n_c}_{i=1}x_i}{n_c}\right)^2 + \left(y_p - \frac{\sum^{n_c}_{i=1}y_i}{n_c}\right)^2.
\end{equation}
Considering the convex function $\Psi(x) = (C-x)^2$ for any real number $C$ and applying Jensen's inequality, we get 
\begin{equation}
    d_{\text{2d}}^2(\mathcal{C}_{c,P}, \bar{P}_{\mathcal{C}_c}) \leq \frac{1}{n_c} \sum^{n_c}_{i=1} (x_p - x_i)^2 + (y_p - y_i)^2.
\end{equation}
From \eqref{eqn:max_distance_pts}, the summation term $(x_p - x_i)^2 + (y_p - y_i)^2$ cannot exceed $R^2_\text{A}$. Hence, we derive
\begin{equation}
     d_{\text{2d}}^2(\mathcal{C}_{c,P}, \bar{P}_{\mathcal{C}_c}) \leq R^2_\text{A},
\end{equation}
completing the proof.
\end{proof}
Given the established theorem, to ensure that all GNs within a DBS cluster are no further than a distance of $R_\text{A}$ from the DBS, it is necessary for the two clusters $\mathcal{C}_a$ and $\mathcal{C}_b$ being merged to meet condition \eqref{eqn:hierc_clus_cond_2}. In conclusion, 
the subset of cluster pairs to be considered for merging at iteration \( t \) is given by
\begin{equation}\label{eqn:hierc_subset}
\mathcal{P}^t = \left\{ (\mathcal{C}_a, \mathcal{C}_b) : \eqref{eqn:hierc_clus_cond_1} \text{ and } \eqref{eqn:hierc_clus_cond_2} \right\}.
\end{equation}
Algorithm~\ref{alg} outlines the pseudocode for our proposed HC method. At each iteration \( t \), two clusters are combined to form a single cluster. Conventionally, a linkage matrix of size \( U \times 4 \) is utilized. The first two columns of row \( t \) in this matrix store the indices of the clusters merged at iteration \( t \). Meanwhile, the third and fourth columns record their dissimilarity value and the total number of data points in the new cluster, respectively. This matrix then facilitates the final clustering determination.

\begin{algorithm}[!ht]
 \caption{DBSs Placement}
 \begin{algorithmic}[1]\label{alg}
 \renewcommand{\algorithmicrequire}{\textbf{Input:}}
 \renewcommand{\algorithmicensure}{\textbf{Output:}}
 \ENSURE $M$, $\boldsymbol{m}_j\; \forall j \in \{1, 2,..,M\}$
 \REQUIRE  $\boldsymbol  u_n\; \forall n \in \{1, 2,..,U\}$, $d_\text{max}$, $N_B$ 
 \\ \textit{Initialisation}: Begin with \( U \) clusters at GNs locations. Using \eqref{eqn:dissimilarity_matrix}, \eqref{eqn:degrees_matrix}, \eqref{eqn:degrees_vector}, and \eqref{eqn:hierc_subset}, compute \( \boldsymbol{D}^0(\mathcal{U})\), \( \hat{\boldsymbol{D}}^0\), \( \boldsymbol{C}^0\), and \( \mathcal{P}^0 \), respectively.
 \STATE Set \( \boldsymbol{R}^0 \) as a copy of \( \boldsymbol{D}^0 \).
  \WHILE {\( \mathcal{P}^t \) is not empty}
  \STATE Choose clusters \( \mathcal{C}_a \) and \( \mathcal{C}_b \) from \( \mathcal{P}^t \) that meet the condition \eqref{eqn:minimum_distance}.
  \STATE Merge the selected clusters and compute \( \boldsymbol{D}^t \) and \( \boldsymbol{R}^t \) using \eqref{eqn:upgmc_update} and \eqref{eqn:complete_link_update}, respectively.
  \STATE Update, if necessary, to generate \( \hat{\boldsymbol{D}}^t\), \( \boldsymbol{C}^t \), and \( \mathcal{P}^t \).
  \ENDWHILE
 \end{algorithmic}
 \end{algorithm}
\subsection{Complexity of Proposed agglomerative HC Algorithm}
There are $U-t$ clusters at each iteration of the proposed algorithm, and $\binom{U - t}{2} = \frac{(U-t)(U-t-1)}{2}$ possible pairs to be considered for merging in the worst case when $\mathcal{P}^t$ covers the entire set of pairs. Thus, the total number of pairs that have to be examined throughout the whole clustering process is $\sum_{t=0}^{U-1} \binom{U - t}{2} = \sum_{n=1}^{U} \binom{n}{2} = \frac{(U - 1)U(U + 1)}{6}$ implying a complexity of $O(U^3)$. Given that the computation of $\mathcal{P}^t$ and the updates to $\boldsymbol{D}^t$, $\hat{\boldsymbol{D}^t}$, and $\boldsymbol{R}^t$ each iteration are bounded by $O(U^2)$, these operations do not alter the overall complexity of the HC algorithm. However, an $O(U^3)$ complexity may be impractical for large $U$. A common solution is \textit{preclustering} to condense initial elements, thereby reducing $U$'s magnitude \cite{hierc_clustering}. Such a strategy is reasonable when deploying DBSs, as GNs in proximity can be aggregated into a single representative location, enhancing the efficiency of our HC method. Now that we have presented the clustering algorithm which determines the locations of the DBSs, the next step is to select their backhaul interconnections using the following GA.

\section{Genetic Algorithm for Backhaul Optimization}\label{sec:ga}
We present a GA which is a type of metaheuristic algorithms for solving problems where traditional methods falter due to the vast search space or problem non-linearity. We begin by encoding potential solutions as chromosomes that are strings of data that represent different instances of the problem. The initial step involves encoding potential solutions as chromosomes—data strings representing problem instances. The effectiveness of each chromosome is assessed through a fitness function, a key quantitative indicator of solution quality. The fitness function plays a crucial role in guiding the evolutionary process towards the most promising search space regions. The GA employs three main operators that mimic natural genetic processes \cite{michalewicz1996evolutionary}: 
\begin{enumerate}
    \item \textit{Selection} acts as a filter, preferring chromosomes of higher fitness for reproduction, thus favoring the transfer of advantageous traits to future generations.
    \item \textit{Crossover} enhances the gene pool's variety by mixing segments from chromosome pairs. Similar to biological recombination, it enables offspring to inherit traits from both parents, potentially yielding superior qualities.
   \item \textit{Mutation}, introducing spontaneous, minor alterations to chromosomes, fosters new variations within the population. This operator expands the search space exploration and aids in circumventing local optima, ensuring diversity and the integrity of the search process.
\end{enumerate}
% In following sections, we explore the details of our GA implementation. We elucidate how solutions are represented, elaborate on the development of a fitness function customized to the unique demands of our problem, and describe the calibration of genetic operators to guarantee a resilient search process. Additionally, we examine the termination criteria and the integration of various heuristics designed to boost the algorithm's efficacy and efficiency. 

\subsection{Representation}
In our GA implementation, each genome comprises concatenated backhaul paths as outlined in section \ref{subsec:dnp}. It's possible for some sequences to consist solely of a single MBS element, indicating unused connections to that MBS. Representing DBS elements within $\mathcal{V}$ by $d_j$ for $j \in \{1,..,M\}$, and MBS elements by $b_i$ for $i \in \{M+1,..,M+B\}$, an example genome might be $\mathcal{S} = \{b_7, d_6, d_1, d_4, d_5, b_8, d_2, d_3, b_9\}$. This configuration highlights an unused MBS ($\mathcal{S}^1$ containing only $b_7$) and two distinct backhaul paths: $\mathcal{S}^2 = \{d_6, d_1, d_4, d_5, b_8\}$ and $\mathcal{S}^3 = \{d_2, d_3, b_9\}$. Genomes are consistently arranged by the MBS index, ensuring $b_{M+B}$ is the final element.

\subsection{Initial Population Generation}
The initial genomes population is generated through permutation of DBSs and random insertion of MBSs. The process is streamlined as follows:

\begin{enumerate}
    \item \textit{DBSs Permutations}: We start by permuting the $M$ DBSs to explore all potential visitation orders. This yields $M!$ unique sequences, each representing a distinct connectivity route among DBSs.
    \item \textit{Random MBS Insertion} Next, we integrate MBSs into the DBSs sequences. Excluding the terminal MBS, we randomly insert each remaining MBS at points ranging from the sequence's start ($0$) to its end ($M$), allowing for repeated insertion points to accommodate scenarios where multiple BSs occupy subsequent position but only the first is actively linked. This results in $(M+1)^{B-1}$ variations.
\end{enumerate}

Therefore, the general number of potential genomes, reflecting the combination of DBS permutations and BS insertion variations, is given by $M! \times (M+1)^{B-1}$. A number of random initial genomes, $N_\text{initial}$, is selected from this population.

\subsection{Mutation}
In our proposed GA the mutation operation involves swapping two random distinct elements within the genome. This ensures that the uniqueness of each element is maintained while still introducing the necessary variability. The frequency of these mutations is carefully controlled to balance between maintaining diversity in the population and preserving promising solutions.

\subsection{Crossover}

Crossover is essential for combining and recombining genetic material from different parent genomes, fostering the creation of new and potentially superior solutions. Our crossover strategy involves selecting complete separate backhaul paths (terminating with a MBS). To ensure the correctness of the offspring individuals, the chains selected correspond to different BS from each parent. We also need to avoid repeated DBS or inappropriate resulting genome length. To address this, we designate one parent as the \textit{leader}, whose chains remain unaltered, and the other as the \textit{follower}. For the follower's contribution, any repetitive elements in the chains are replaced with unused ones. Furthermore, discrepancies in length are adjusted by either removing duplicate elements or adding missing ones to match the overall structure. This method ensures the production of coherent offspring genomes, incorporating a mix of genetic material without duplicating DBSs or distorting the intended genome length.

\subsection{Fitness}
The fitness function is crucial for steering the evolutionary process towards superior solutions by favoring individuals with higher fitness for reproduction. Not all intermediary solutions fully meet the required connectivity and flow conditions, necessitating a scoring system that accurately reflects their effectiveness. For each backhaul link in a given solution $\mathcal{S}$, we assess the residual capacity as
\begin{equation}\label{eqn:fitness_edge_residualcapacity}
p'(\mathcal{S}^p_r,\mathcal{S}^p_{r+1})=R(\mathcal{S}^p_r,\mathcal{S}^p_{r+1})-\sum_{j=1}^{r}{G(j)}.
\end{equation}
A valid solution should satisfy \eqref{eqn:opt0_cond4} by ensuring that 
\begin{equation}\label{eqn:ga_condition}
p'(\mathcal{S}^p_r,\mathcal{S}^p_{r+1}) \geq 0 \;\;\; \forall r \in \{1,..,|\mathcal{S}^p|-1\}, \forall p \in \{1,..,M\}.
\end{equation}
Then, we define two different fitness evaluation functions as
\begin{equation}\label{eqn:fitness_function_edge}
  F_{\text{edge}}(\mathcal{S})=\sum_{p=1}^{B}\sum_{r=1}^{|\mathcal{S}^p|-1}{p'\left(\mathcal{S}^p_r,\mathcal{S}^p_{r+1}\right)},
\end{equation}
and 
\begin{equation}\label{eqn:fitness_function_node}
  F_{\text{node}}(\mathcal{S})=\sum_{p=1}^{B}{\sum_{r=1}^{|\mathcal{S}^p|-1}\min_{k = 1,...,r}p'\left(\mathcal{S}^p_k,\mathcal{S}^p_{k+1}\right)}.
\end{equation}
The $F_{\text{node}}(\mathcal{S})$ function in \eqref{eqn:fitness_function_node} takes into account the fact that in order to increase transfer at a node, there must be adequate flow reserves on all edges leading to the base station. While $F_{\text{edge}}(\mathcal{S})$ in \eqref{eqn:fitness_function_edge} does not account for this aspect and focuses on the potential contribution of individual edges only. However, it has an advantage during the crossover stage of the genetic algorithm. Individuals evaluated with $F_{\text{edge}}(\mathcal{S})$ may possess better potential capabilities for creating offsprings that show more promising prospects.

\subsection{Evaluation of Fitness Functions}
In our work, the $F_{\text{node}}(\mathcal{S})$ is always used to evaluate the obtained final solutions and their validity is ensured by \eqref{eqn:ga_condition}. However, during the GA process, individual genomes are evaluated for reproduction using either $F_{\text{node}}(\mathcal{S})$ or $F_{\text{edge}}(\mathcal{S})$. The results obtained from each case shown in the following section for the sake of presenting a comprehensive comparison. Additionally, we opted to evaluate three distinct penalty methods for the fitness function when selecting individuals for subsequent generations in cases of incorrect solutions: applying no penalty, imposing a penalty equivalent to the insufficient residual capacity as indicated by the $p'$ \eqref{eqn:fitness_edge_residualcapacity} function, and deducting a constant penalty value from the overall function score. This constant penalty value is based on the maximum positive value ensuring that the value with penalty is always negative.

The following six different settings of the fitness function and penalty were studied:
%\footnote{It is crucial to note that the selection of solutions always relies on the node surplus $F_{\text{node}}(\mathcal{S})$ \eqref{eqn:fitness_function_node} function with a correctness condition $p'$ \eqref{eqn:fitness_edge_residualcapacity}.}
\begin{enumerate}
\item Using edge surplus without penalty (ENP).
\item Using edge surplus with a penalty value (EVP).
\item Using edge surplus with a penalty corresponding to deficits on edge flows (EEP).
\item Using node surplus without penalty (NNP).
\item Using node surplus with a penalty value (NVP).
\item Using node surplus with a penalty corresponding to deficits on edge flows (NEP).
\end{enumerate}

\subsection{Algorithm}

Algorithm~\ref{alg:genetic} outlines the pseudocode for our GA. This algorithm is distinctive in its use of two separate fitness functions: one for selecting the solution ($F_{\text{sol}}$) and another for selecting individuals for the next population ($F_{\text{sel}}$). This dual fitness function approach is unusual in genetic algorithms and is tailored to enhance both the quality of solutions and the certainty of finding a solution, considering that not all individuals meet the mandatory requirements. The stopping condition for the GA is achieved by reaching the maximum number of generations. 

\begin{algorithm}[!ht]
 \caption{Genetic Algorithm for Backhaul Optimization}
 \begin{algorithmic}[1]\label{alg:genetic}
 \renewcommand{\algorithmicrequire}{\textbf{Input:}}
 \renewcommand{\algorithmicensure}{\textbf{Output:}}
 \REQUIRE Population size $N$, stop condition $SC$, mutation rate $\chi_\text{M}$, crossover rate $\chi_\text{CO}$, fitness selection function $F_{\text{sel}}$, fitness solution function $F_{\text{sol}}$
 \ENSURE Optimized solution
 \\ \textit{Initialisation}: Generate initial population of $N_\text{initial}$ individuals.
  \REPEAT
  \STATE Evaluate fitness function (\eqref{eqn:fitness_function_node} and \eqref{eqn:fitness_function_edge}) of each individual.
  \STATE Store individual with the best solution fitness $F_{\text{sol}}$. 
  \STATE Select parents basing on selection fitness $F_{\text{sel}}$. 
  \STATE Generate new population using crossover of parents with rate $\chi_\text{CO}$.
  \STATE Apply mutation to new population with a rate of $\chi_\text{M}$.
  \STATE Replace the old population with the new population.
  \STATE Perform elitism by preserving the top $\chi_\text{E}$ individuals.
  \UNTIL {$SC$}
 \STATE Return the best stored individual.
 \end{algorithmic}
\end{algorithm}

\section{Simulation and Results}\label{sec:sim}
For every simulation run, we create a fresh distribution of GNs based on the PCP outlined in \cite{pcp_user_modeling}. In this approach, each point from a Poisson Point Process (PPP) serves as a nucleus for forming a cluster of GNs in its vicinity. The simulations are run in Python and for the GA algorithm we leverage PyGAD \cite{gad2023pygad}. The MBSs are located at the four corners of the area. Simulation parameters are listed in Table~\ref{tab:params}.
%TwoColumn
\begin{table}[htbp]
\begin{center}
\caption{Summary of Key Simulation Parameters}
\label{tab:params}
\begin{tabular}{||p{0.6\linewidth}|p{0.26\linewidth}||}
\hline  
\textbf{Parameter} & \textbf{Value} \\
\hline
\hline  
DBS height ($h$), environment variables ($\alpha, \beta$)  & 60 m, $9.61$, $0.16$\\
\hline 
Expected PL ($\eta_{\text{LOS}}$, $\eta_{\text{NLOS}}$), carrier frequency ($f_c$), GN load ($R_n$), MBS count ($B$) & 1 dB, 20 dB, 2 Ghz, 20 Mbps, 4 \\
\hline  
Beam waist radius ($\omega_0$),  weather coefficient ($\kappa$), FSO Power ($P_\text{FSO}$) & 0.25 cm, $4.3\times 10 ^{-4}$ $m^{-1}$, 50 mW\\
\hline  
Wavelength ($\lambda$), lens radius ($r_0$), Responsivity ($\eta$), noise Power ($\sigma_n^2$) & 1550 nm, 0.1 m, 0.5, -60.1 dBm \\
\hline  
GA's generations maximum count, initial population size, Crossover rate ($\chi_\text{CO}$), mutation rate ($\chi_\text{M}$), elitism rate($\chi_\text{E}$) & 400, 400, 0.3, 0.2, 0.1\\
\hline  
\end{tabular}
\end{center}
\end{table}
% \begin{table}[htbp]
% \begin{center}
% \caption{Summary of Key Simulation Parameters}
% \label{tab:params}
% \begin{tabular}{||p{0.7\linewidth}|p{0.28\linewidth}||}
% \hline  
% \textbf{Parameter} & \textbf{Value} \\
% \hline
% \hline  
% DBS height ($h$), environment variables ($\alpha, \beta$)  & 60 m, $9.61$, $0.16$\\
% \hline 
% Expected PL ($\eta_{\text{LOS}}$, $\eta_{\text{NLOS}}$), carrier frequency ($f_c$) & 1 dB, 20 dB, 2 Ghz \\
% \hline  
% Coverage radius ($ R_\text{A}$), beam waist radius ($\omega_0$),  weather coefficient ($\kappa$) & 200 m, 0.25 cm, $4.3\times 10 ^{-4}$ $m^{-1}$\\
% \hline  
% Wavelength ($\lambda$), lens radius ($r_0$), Responsivity ($\eta$) & 1550 nm, 0.1 m, 0.5 \\
% \hline  
% FSO Power ($P_\text{FSO}$), noise Power ($\sigma_n^2$), GN load ($R_n$) & 50 mW, -60.1 dBm, 20 Mbps \\
% \hline  
% MBS count ($B$), GA's generations maximum count, GA's initial population size & 4, 400, 400\\
% \hline
% GA's crossover rate ($\chi_\text{CO}$), mutation rate ($\chi_\text{M}$), elitism rate ($\chi_\text{E}$) & 0.3, 0.2, 0.1\\
% \hline  
% \end{tabular}
% \end{center}
% \end{table}
\subsection{Performance of Proposed HC Algorithm}
\begin{figure}[!htb]
  \centering
  \begin{minipage}{0.48\textwidth}
\centering
  \includegraphics[width=1\textwidth, trim={0.3cm 0.0cm 0cm 0.5cm},clip]{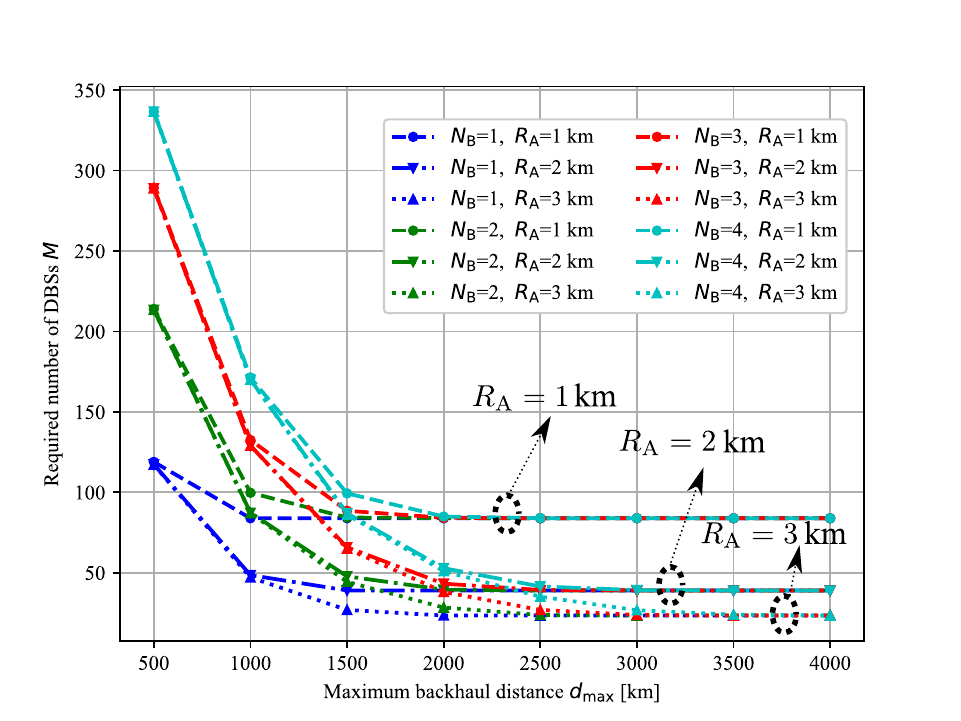}
  \caption{Average minimum required number of DBSs for different backhaul distances $d_\text{max}$ given the specific $N_\text{B}$ and $R_\text{A}$ constraints.}
  \label{fig:n_dbs_vs_fso}
  \end{minipage}\hfill
  \begin{minipage}{0.48\textwidth}
   \centering
  \includegraphics[width=1\textwidth, trim={0.3cm 0.5cm 0cm 0.0cm},clip]{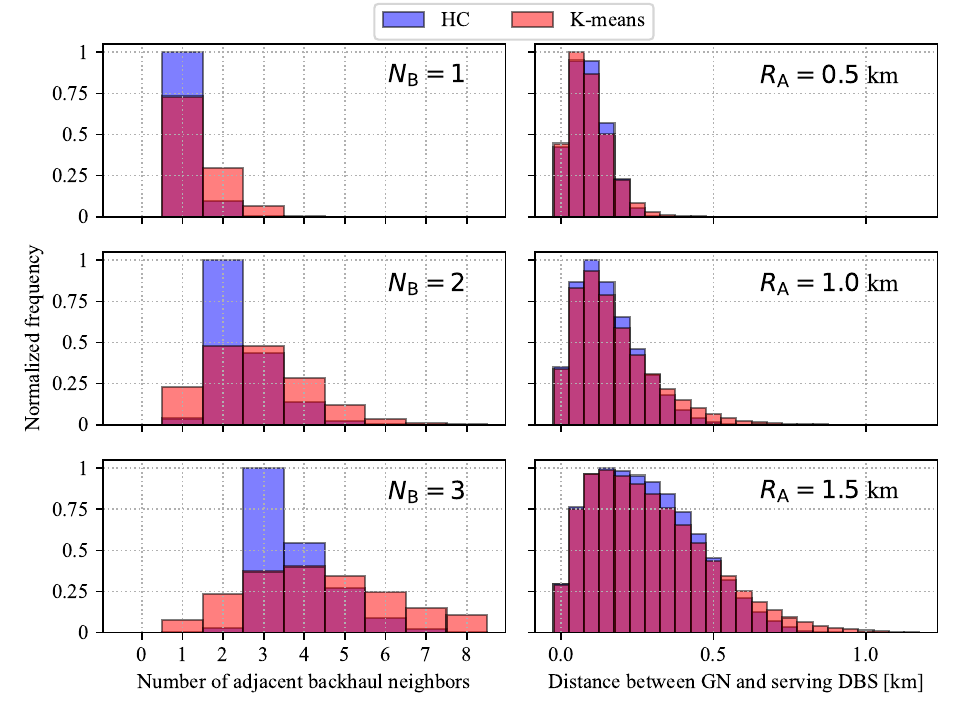}
\caption{Left: Histogram of DBS backhaul neighbors across $N_\text{B}$ settings. Right: Histogram of GN-DBS distances for various $R_\text{A}$ settings. Results from the proposed HC method are compared with K-means, keeping the cluster count ($M$) constant.}
  \label{fig:km_vs_hc}
  \end{minipage}
  \end{figure}

To obtain the results below we simulated the HC algorithm under varying $N_\text{B}$ and $R_\text{A}$ values. Fig.~\ref{fig:n_dbs_vs_fso} illustrates the required number of DBSs to fulfill the constraints, derived from the resulting cluster count, $U-t$, when $\mathcal{P}^t$ is exhausted at iteration $t$ (see Algorithm~\ref{alg}). It is evident that with relatively low values of $d_\text{max}$, the number of DBSs needed increases non-linearly with the $N_\text{B}$ parameter which emphasizes the expected trade-off between backhaul reliability and DBSs count when $d_\text{max}$ is small. However, as $d_\text{max}$ grows larger, the effect of $N_\text{B}$ diminishes, leaving the disk radius $R_\text{A}$ as the predominant factor that seems to have more drastic effect on $M$. This influence can be effectively modulated through adjusting DBSs height $h$ (see \eqref{eqn:expectedPathLoss})\cite{optimalLapAltitude}. In Fig.~\ref{fig:km_vs_hc}, we evaluate the HC clustering against K-means clustering with K-means++ initialization \cite{kmeans++} by examining DBS backhaul neighbor counts and GN-DBS distances. For neighbor analysis, we set $R_\text{A} = \infty$ in HC to omit the fronthaul constraint, and for distance analysis, we set $d_\text{max} = \infty$. Unlike K-means, HC can adhere to these constraints. Both methods use the same number of DBSs ($M$), determined after depleting $\mathcal{P}^t$ during HC as previously described. The left figures show neighbor counts for various $N_\text{B}$ settings with $d_\text{max} = 1$ km. At $N_\text{B}=2$, HC nearly meets the neighbor constraint for all DBSs besides few clusters that are resulting from outliers (i.e., GNs that are far away from the remaining points). Nevertheless, HC surpasses K-means in maintaining neighbor counts. On the right, GN-DBS distance distributions are similar for both algorithms, but K-means' distribution tails longer, often breaching the $R_\text{A}$ limit, whereas HC consistently respects it.

\begin{figure}[!htb]

\end{figure}

% Examples of two extreme test instances can be found in Fig.~\ref{fig:example_net}.

% \begin{figure}
%     \begin{minipage}{0.45\textwidth}
%         \centering
%         \includegraphics[width=\linewidth]{figures/examle_net_i1d10r2000_in.png}
%         \small{The lack of sufficient backhaul links renders the problem unsolvable.\vspace{12mm}}
%     \end{minipage}
%     \hfill
%     \begin{minipage}{0.45\textwidth}
%         \centering
%         \includegraphics[width=\linewidth]{figures/examle_net_i5d49r4000_sg.png}
%         \small{Each base station is represented by a distinct color, and the colored edges depict an example solution of the communication network created among the drones.}
%     \end{minipage}
%     \caption{Visualization of the extremes: the smallest and the largest test instances by drone Number and communication range.}
%     \label{fig:example_net}
% \end{figure}

\subsection{Comparison of different GA settings}
In this section we show the performance of the proposed GA with different settings (i.e., selected fitness function and penalty strategy) for a limited number of iterations, compared with a benchline exact algorithm that exhaustively searches a random number of $N_\text{exact} = 10^7$ of possible solutions from the solution space. The results are obtained for $R_\text{A}= \infty$\footnote{We set $R_\text{A}= \infty$ to neglect the access layer and to analyze the performance of the GA considering only the factors that directly relate to the backhaul layer} and $N_\text{B} =2$.
%, and when unspecified, $d_\text{max} =????$ and $M=????$. 
\subsubsection{Probability of finding a correct solution}
\begin{figure}[!htb]
  \centering
  \begin{minipage}{0.45\textwidth}
    \centering
    \includegraphics[width=1\linewidth]{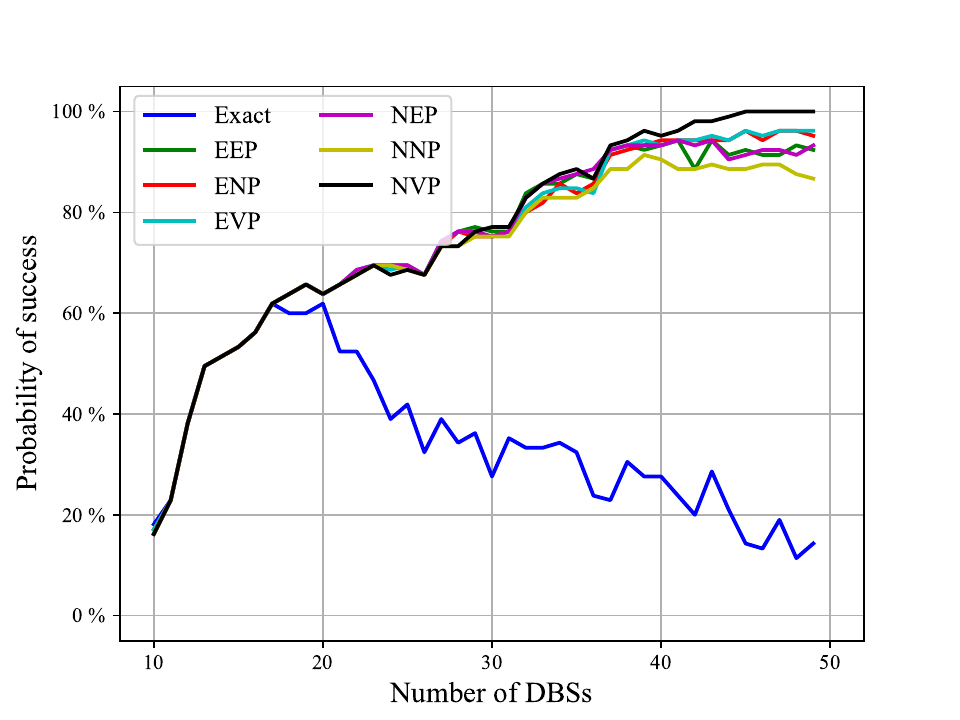}
    \caption{Probability of success vs. DBSs count $M$ for different GA settings and the benchline scheme. The NVP strategy outperforms the rest for large values of $M$.}
    \label{fig:SuccessProbabilityByNodes}
  \end{minipage}\hfill
  \begin{minipage}{0.45\textwidth}
    \centering
    \includegraphics[width=1\linewidth]{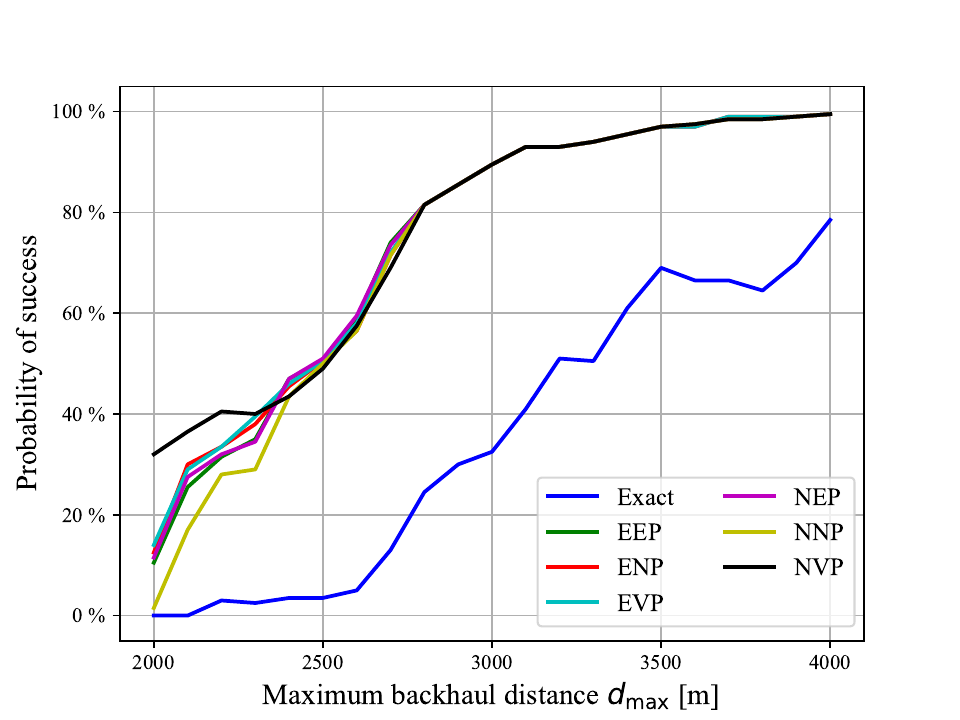}
    \caption{Probability of success vs. backhaul distance $d_\text{max}$ for different GA settings and the benchline scheme. For small backhaul distances NVP outperforms other settings.}
    \label{fig:SuccessProbabilityByDistance}
  \end{minipage}
  \end{figure}

Fig~\ref{fig:SuccessProbabilityByNodes} plots the probability of finding a correct solution before reaching the stopping condition for different values of DBSs count $M$. As shown, for a small number of DBSs, the results do not significantly differ between genetic algorithms and the exact algorithm. However, beyond this number, the exact algorithm performs notably worse, which is not surprising given the NP-complete nature of the problem. From around 30 drones up to 40, increasing differences between the various versions of fitness functions become apparent, stabilizing above 40 drones with the best and worst differing by about 13\% – the worst version (NNP) yielding around 87\% success rate, and the best (NVP) reaching 100\%. Similarly, Fig.~\ref{fig:SuccessProbabilityByNodes} plots the success probability for different backhaul distance values. We observe that the most challenging cases occur with the shortest $d_\text{max}$ constraint. It is evident that above a distance of 2.4 km, the results vary little, and above 2.8 km, they are identical for all genetic algorithms. Naturally, this applies only to genetic algorithms, as the exact algorithm performs considerably worse due to its inability to handle larger instances. Below the aforementioned 2.4 km threshold, a dominance of the NVP version is observed, similar to the previous analysis. In the extreme case of a 2 km range, NVP produces results for three times more instances than other versions of genetic algorithms, except for the NNP version, which performs surprisingly poorly, on par with the exact algorithm.

\subsubsection{Quality of achieved correct solutions}
\begin{figure}[!htb]
  \centering
  \begin{minipage}{0.45\textwidth}
    \centering
    \includegraphics[width=1\linewidth]{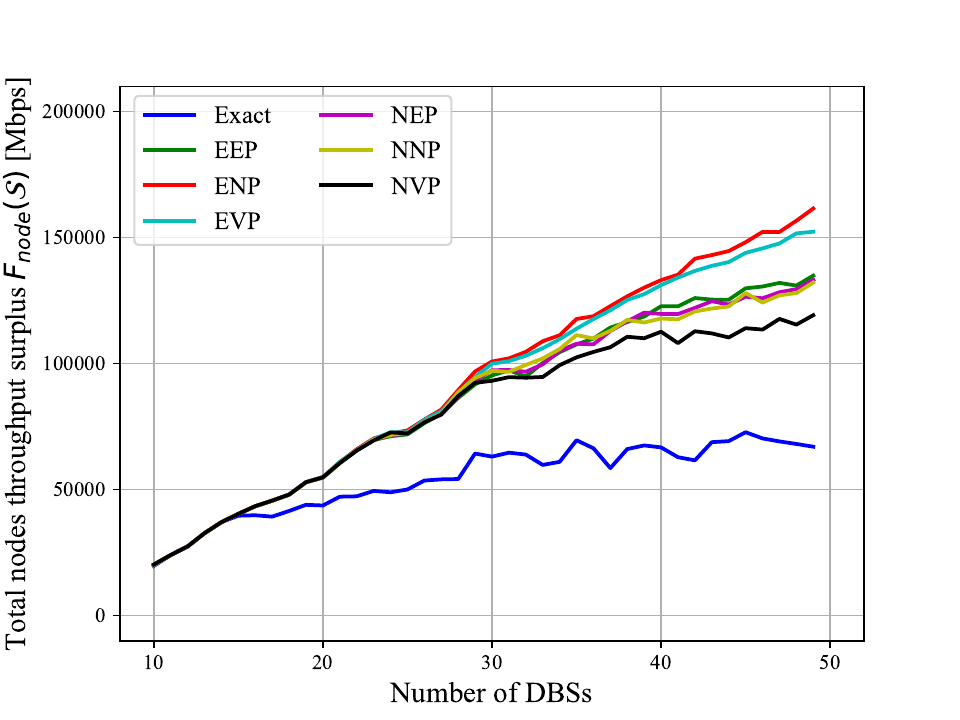}
    \caption{Total nodes throughput surplus $F_{\text{node}}(\mathcal{S})$ vs. DBSs count $M$ for different GA settings and the benchline scheme. As $M$ grows larger, the ENP strategy performs the best while the NVP achieves the least surplus.}
    \label{fig:ScoreByNodes}
  \end{minipage}\hfill
  \begin{minipage}{0.45\textwidth}
    \centering
    \includegraphics[width=1\linewidth]{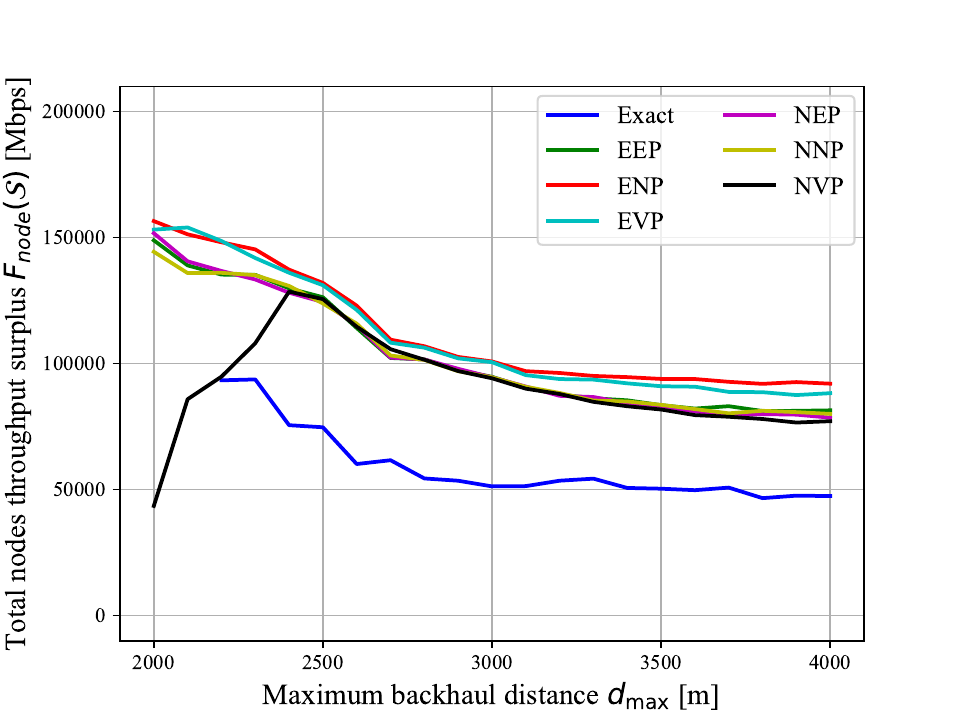}
    \caption{Total nodes throughput surplus $F_{\text{node}}(\mathcal{S})$ vs. backhaul distance $d_\text{max}$ for different GA settings and the benchline scheme. Again, the ENP strategy excels while the NVP achieves the least surplus.}
    \label{fig:ScoreByDistance}
  \end{minipage}
\end{figure}

Now we compare the quality of different correct solutions. The quality is assessed through the sum of the transmission bandwidths that can be increased on drones, a concept described by the function $F_{\text{node}}(\mathcal{S})$ \eqref{eqn:fitness_function_node}. This measure is deemed appropriate as it ensures the highest possible 'reserve' to accommodate any increase in user needs. Importantly, this reserve helps to avoid the necessity of increasing the number of deployed DBSs or FSO transmission power when the required rate of GNs increases. Similarly to before, the results are presented in two distinct ways: Fig.~\ref{fig:ScoreByNodes} shows the outcomes based on the number of drones, while Fig.~\ref{fig:ScoreByDistance} illustrates the results with respect to the maximum backhaul distance $d_\text{max}$. We observe in Fig.~\ref{fig:ScoreByNodes} that the quality of solutions consistently increase, which aligns with expectations since the outcome measures the cumulative transmission reserve for all drones. Up to 15 drones, the results are identical, likely due to the ease of finding the best solution for such small instances. However, beyond this count and up to about 30, genetic algorithms tend to offer similar results, while the exact algorithm struggles to keep pace in the limited time frame. Beyond 30 drones, we notice a divergence in the results; the ENP version excels, maintaining high-quality outcomes despite the increasing number of drones, whereas the NVP version shows the least favorable results. Similar to the probability analysis, the chart examining outcome quality by distance (Fig.~\ref{fig:ScoreByDistance}) shows the greatest variation in the most challenging cases, i.e., short communication ranges. The results for genetic algorithms are relatively close, with variations not exceeding 20\%, but the exact algorithm yields outcomes approximately half as good. Notably, the ENP algorithm performs the best in this scenario. In this case, NVP ranks as the least effective, especially in scenarios with limited communication range. 

The NVP version's comparatively lower quality of outcomes for communication distances under 2.4 km appears to be closely linked to its ability to find solutions in a greater number of cases. For instance, at a distance of 2 km, the NVP version finds a solution in 64 out of 200 instances, whereas the ENP version succeeds in only 25 out of 200 instances. Further analysis of the cases shows that the scenarios where NVP had an advantage were those where the solutions scored the lowest. This implies that the NVP fitness function excels at finding solutions that marginally exceed the required threshold. From this observation, we infer that the NVP version of the fitness function is particularly effective at identifying solutions, even those that only slightly surpass the necessary criteria. This makes it a robust choice in scenarios where securing any viable solution is prioritized, especially in the most challenging situations with limited interconnectivity of DBSs.

\subsection{GA Probability of Success with HC vs. K-means}
\begin{figure}[!htb]
\centering
  \includegraphics[width=0.45\textwidth, trim={0.3cm 0.1cm 1cm 0.8cm},clip]{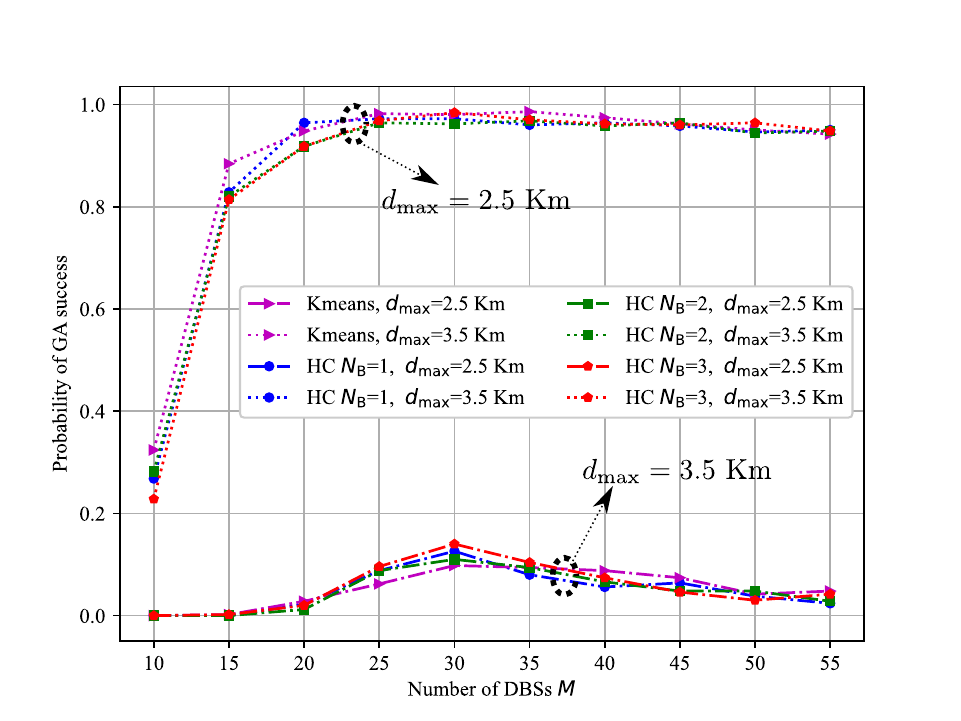}
  \caption{Probability of GA success vs. number of DBSs, $M$, for different $d_\text{max}$ values. DBSs are either placed using HC with different $N_\text{B}$ values while setting $R_\text{A}= \infty$, or according to K-means.}
  \label{fig:solution_probability}
\end{figure}

In Fig~\ref{fig:solution_probability}, we compare the performance of the proposed NVP version of the GA assuming two different DBSs placement strategies. As a benchline strategy we again employ K-means, and we compare the obtained results with the cases where our proposed HC algorithm is employed assuming different $N_\text{B}$ settings. Since there is no way for K-means to conserve the coverage radius constraint, we set $R_\text{A}= \infty$. The performance of both strategies is largely similar. With a relaxed $d_\text{max}$ of $3.5$~km, K-means seems to be performing slightly better with a small $M$. This occurs because K-means achieves less uniform distribution than HC (see Fig.~\ref{fig:km_vs_hc}), which shrinks the solution search space, allowing for better solutions to be found earlier. On the other hand, for a more stringent $d_\text{max}$ of $2.5$~km, both strategies achieve similar results. Also, we observe that for the specified limited number of generations and initial population size, increasing the number of DBSs decreases the probability of success, as the solution space size increases. Therefore, we conclude that our proposed HC algorithm is able to conserve the coverage disk and backhaul neighbors constraint without degrading the performance of the GA when compared with the well-known K-means algorithm.

\section{Conclusion}\label{sec:conclusion}
Considering a complex network structure with aerial platforms of various types deployed to provide wireless connectivity and relay backhaul traffic, we formulated and addressed the problem of simultaneously placing DBSs and optimizing their backhaul mesh relay network. This challenge was tackled by dividing it into two subproblems: positioning DBSs to ensure users are within their communication range and that each DBS is within reach of at least $N_\text{B}$ backhaul neighbors; and identifying ptp links to establish, aiming for the highest possible margin in satisfying the load across each backhaul path. We proposed an HC algorithm and a GA to solve these problems efficiently. The proposed HC was shown to satisfy the coverage disk constraint and almost completely ensure the number of backhaul adjacency constraint except in the cases of outliers DBSs that are far from other clusters. Moreover, the proposed GA was shown be able to find solutions relatively fast, which makes it a viable solution for the novel DNP problem which, to the best of the authors' knowledge, is not yet examined in the literature. Finally, it was found that the proposed HC performs similarly the benchline K-means algorithm, whereas the latter cannot conserve the coverage disk and backhaul adjacency constraints.

% \section*{Acknowledgments}
% This should be a simple paragraph before the References to thank those individuals and institutions who have supported your work on this article.

\bibliography{bibtex/bib/IEEEabrv.bib,bibtex/bib/references.bib}{}
\bibliographystyle{IEEEtran}
\newpage

 

% \begin{IEEEbiography}[{\includegraphics[width=1in,height=1.25in,clip,keepaspectratio]{figures/3d.png}}]{Michael Shell}
% \end{IEEEbiography}



\vfill

\end{document}